\documentclass{aa}  
\usepackage{supertabular,booktabs}
\usepackage{graphicx}
\usepackage{txfonts}
\usepackage{comment}
\usepackage{xcolor}
\usepackage[colorlinks,citecolor=blue]{hyperref}
\newcommand{\program}[1]{\texttt{#1}}

\begin{document} 

   \title{SN\,2020qlb: A hydrogen-poor superluminous supernova with well-characterized light curve undulations}

   \subtitle{}

   \author{S.~L.~West\inst{1}
        \and R.~Lunnan\inst{1}
        \and C.~M.~B.~Omand\inst{1}
        \and T.~Kangas\inst{2}
        \and S.~Schulze\inst{3}
        \and N.~L.~Strotjohann\inst{4}
        \and S.~Yang\inst{1}
        \and C.~Fransson\inst{1}
        \and J.~Sollerman\inst{1}
        \and D.~Perley\inst{5}
        \and L.~Yan\inst{6}
        \and T.-W.~Chen\inst{1}
        \and Z.~H.~Chen\inst{7}
        \and K.~Taggart\inst{5,8}
        \and C.~Fremling\inst{6,9}
        \and J.~S.~Bloom\inst{10,11}
        \and A.~Drake\inst{9}
        \and M.~J.~Graham\inst{9}
        \and M.~M.~Kasliwal\inst{9}
        \and R.~Laher\inst{12}
        \and M.~S.~Medford\inst{10,11}
        \and J.~D.~Neill\inst{9}
        \and R.~Riddle\inst{6}
        \and D.~Shupe\inst{12}
          }

   \institute{The Oskar Klein Centre, Department of              Astronomy, Stockholm University,
            AlbaNova, SE-106 91 Stockholm, Sweden\
        \and
            The Oskar Klein Centre, Department of Physics, KTH Royal Institute of Technology, AlbaNova, SE-106 91 Stockholm, Sweden
        \and
            The Oskar Klein Centre, Department of Physics, Stockholm University, AlbaNova, SE-106 91 Stockholm, Sweden
        \and
            Benoziyo Center for Astrophysics, The Weizmann Institute of Science, Rehovot 76100, Israel\
        \and
            Astrophysics Research Institute, Liverpool John Moores University, Liverpool Science Park, 146 Brownlow Hill, Liverpool L35RF, UK\
        \and
            The Caltech Optical Observatories, California Institute of Technology, Pasadena, CA 91125, USA
        \and
            Physics Department and Tsinghua Center for Astrophysics (THCA), Tsingua University, Beijing 100084 China
        \and
            Department of Astronomy and Astrophysics, University of California, Santa Cruz, CA 95064, USA
        \and
            Division of Physics, Mathematics, and Astronomy, California Institute of Technology, Pasadena, CA 91125, USA
        \and
            Department of Astronomy, University of California, Berkeley, Berkeley, CA 94720, USA
        \and
            Lawrence Berkeley National Laboratory, 1 Cyclotron Rd., Berkeley, CA 94720, USA
        \and
            IPAC, California Institute of Technology, 1200 E. California
             Blvd, Pasadena, CA 91125, USA
            }

   \date{Received Xxxxxxx xx, 2022; accepted Xxxx xx, 2022}

 
  \abstract
   {SN\,2020qlb (ZTF20abobpcb) is a hydrogen-poor superluminous supernova (SLSN-I) that is among the most luminous (maximum M$_{g} = -22.25$ mag) and that has one of the longest rise times (77 days from explosion to maximum). We estimate the total radiated energy to be $>2.1\times10^{51}$ erg. SN\,2020qlb has a well-sampled light curve that exhibits clear near and post peak undulations, a phenomenon seen in other SLSNe, whose physical origin is still unknown.}
   {We discuss the potential power source of this immense explosion as well as the mechanisms behind its observed light curve undulations.}
   {We analyze photospheric spectra and compare them to other SLSNe-I. We constructed the bolometric light curve using photometry from a large data set of observations from the Zwicky Transient Facility (ZTF), Liverpool Telescope (LT), and Neil Gehrels Swift Observatory and compare it with radioactive, circumstellar interaction and magnetar models. Model residuals and light curve polynomial fit residuals are analyzed to estimate the undulation timescale and amplitude. We also determine host galaxy properties based on imaging and spectroscopy data, including a detection of the [O III]$\lambda$4363, auroral line, allowing for a direct metallicity measurement.}
   {We rule out the Arnett $^{56}$Ni decay model for SN\,2020qlb's light curve due to unphysical parameter results. Our most favored power source is the magnetic dipole spin-down energy deposition of a magnetar. Two to three near peak oscillations, intriguingly similar to those of SN\,2015bn, were found in the magnetar model residuals with a timescale of $32\pm6$ days and an amplitude of 6$\%$ of peak luminosity. We rule out centrally located undulation sources due to timescale considerations; and we favor the result of ejecta interactions with circumstellar material (CSM) density fluctuations as the source of the undulations.}
   {}
   

   \maketitle
%

\section{Introduction} \label{Intro}

The most luminous of all supernovae (SNe) are the superluminous SNe (SLSNe). Initial findings of these remarkable SNe were first made in the late 1990s but were largely explained away as scaled up versions of known SNe or as SNe Type IIn \citep{Howell2017}. \citet{RichardsonEtal2002} identified a population of rare and overluminous events. Discoveries of relatively nearby superluminous events by, for example, \citet{QuimbyEtal2007}, \citet{SmithEtal2007}, \citet{Gal-YamEtal2009}, and others marked the beginning of intensive study. \citet{Gal-Yam2012} reviewed SLSNe and argued to divide them into subcategories, for example Type II (H-rich) and Type I (H-poor). \citet{Gal-Yam2019} pointed out that, based on \citet{DeCiaEtal2018} and \citet{QuimbyEtal2018}, H-poor SNe with peak luminosities brighter than M$_{g}=-19.8$ mag are spectroscopically similar, wherein the most important connecting features are the O\,II absorption lines. Further details are discussed in recent review articles \citep{Howell2017,Gal-Yam2019,Chen2021,Nicholl2021}.

In general, SLSNe light curves cannot be explained by hydrogen (H) recombination or the decay of typical amounts of $^{56}$Ni, which power the majority of normal SN light curves, thereby suggesting the use of more exotic mechanisms. One such possibility is the pair-instability SN mechanism \citep{HegerWoosley2002} wherein high energy photons could interact with core nucleons to form positron and electron pairs initiating a core-collapse SN explosion and creating the required large amounts of $^{56}$Ni to power the SLSN light curves. A second hypothetical light curve power source is the interaction of the SN ejecta with a circumstellar medium \citep[CSM; e.g.][]{ChatzopoulosEtal2012,SorokinaEtal2016,WheelerEtal2017}. A third power source, the spin-down of a millisecond magnetar, \citep{KasenBildsten2010,Woosley2010} has emerged as a model that can fit the general shape of SLSN-I light curves \citep{InserraEtal2013,NichollEtal2013, NichollEtal2017}.

While the magnetar model has proven successful in reproducing the overall timescales and energetics of SLSNe-I, a significant number of objects also show light curve undulations that are not easily explainable in a simple magnetar spin-down scenario (e.g., \citealt{HosseinzadehEtal2021}). The systematic monitoring and regular cadence of surveys such as the Zwicky Transient Facility (ZTF; \citealt{GrahamEtal2019,BellmEtal2019,MasciEtal2019}) has demonstrated that such undulations are quite common, and are present in as much as $34-62 \%$ of SLSNe-I \citep{ChenEtal2022b}. Similarly, \citet{HosseinzadehEtal2021} found that $44-76 \%$ of SLSNe-I could not be explained by only a smooth magnetar model. The physical mechanism responsible for these undulations or ``bumps'' in the light curves is still an open question -- while simple arguments based on diffusion timescales as well as bump appearance and duration times can place some constraints on them, detailed studies of such light curve undulations in multiple filters are currently lacking. 

Here, we present SN\,2020qlb (ZTF20abobpcb), a luminous and slow rising SLSN-I with prominent near and post peak light curve undulations. Its well-sampled light curve coverage from ZTF, the Liverpool Telescope (LT), and \textit{Swift} enables us to analyze both its primary power source and the nature of the light curve undulations in detail.

This paper is organized as follows. In Sect.~\ref{Data} we present the available photometric and spectroscopic data. In Sections \ref{light-curveProperties} and \ref{SpectralProperties}, we analyze light curve and spectral properties. In Sect.~\ref{Blackbodyfits} we use blackbody fits to estimate the photospheric temperature and radius evolutions. In Sections \ref{Bololight-curve} and \ref{Models}, we discuss the construction of the bolometric light curve and its fit with two relevant power source models. In Sect.~\ref{Undulations} we analyze the model residuals to characterize the apparent light curve undulations. We discuss the host galaxy properties in Section~\ref{sec:host}. In Sect.~\ref{Discussion} we discuss the results relative to current understandings and in Sect.~\ref{Conclusions} we conclude by summarizing our findings. We assume a flat Lambda cold dark matter ($\Lambda$CDM) cosmology with H$_{0}=70$ km~s$^{-1}$ Mpc$^{-1}$, $\Omega_{M}=0.27$, and $\Omega_{\Lambda}=0.73$ \citep{KomatsuEtal2011}. All magnitudes herein are in the AB system \citep{OkeGunn1983}. UT dates are used throughout.

\section{Observations} \label{Data}
\begin{figure}
        \centering
        \includegraphics[width=\columnwidth]{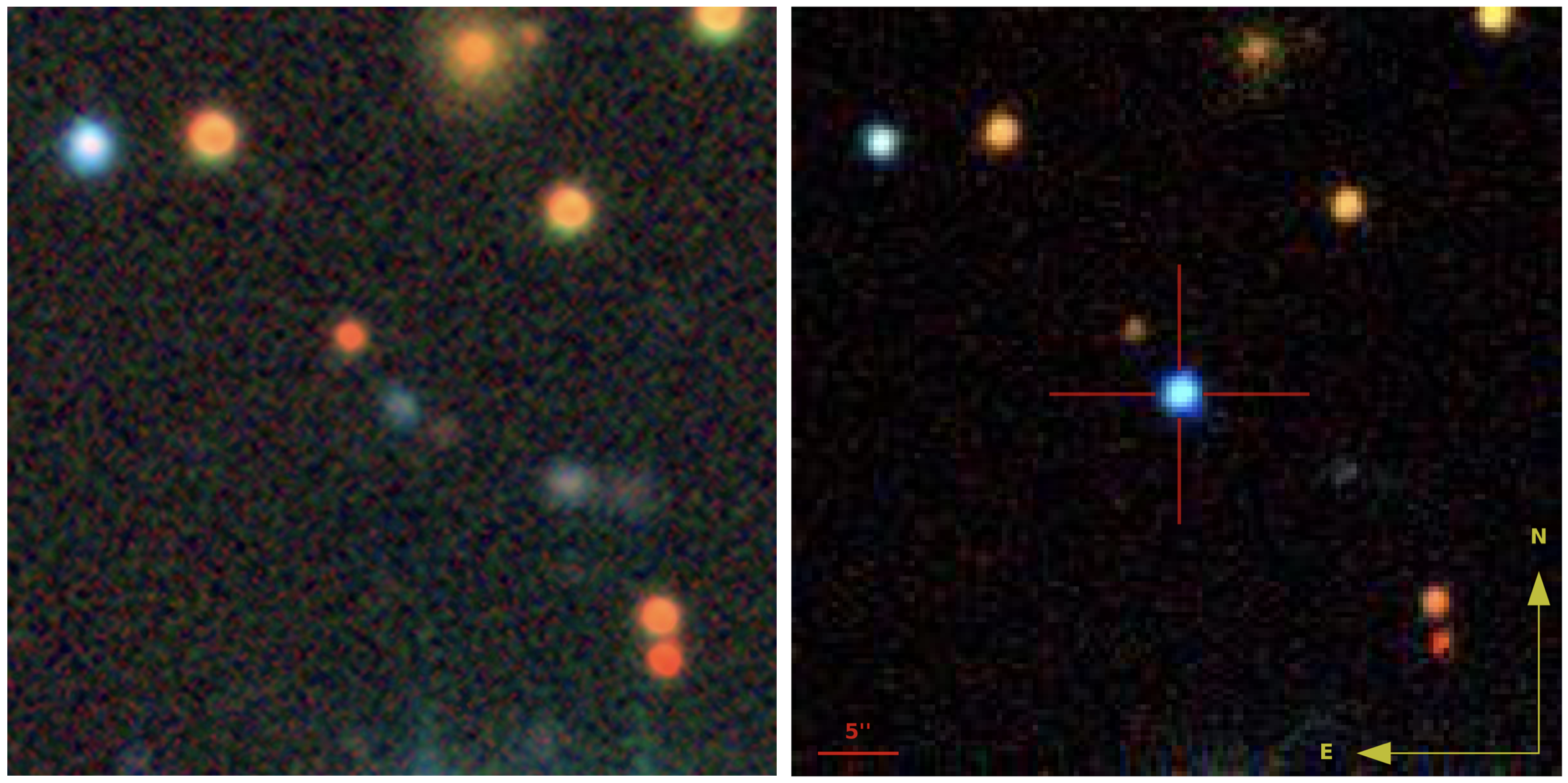}
        \caption{\label{HostGalaxy_Image}Left: Pre-SN image retrieved from the Dark Energy Camera Legacy Surveys (DR9). Right: Image of SN\,2020qlb taken on August 9, 2020 (61 rest frame days before peak) with the Liverpool Telescope in $ugriz$ frames and stacked \citep{LuptonEtal2004} around the SN location. Each image is 45 by 45 arcseconds.}
\end{figure}

SN\,2020qlb was discovered by the ZTF as ZTF20abobpcb on July 23, 2020 at position (J2000) right ascension (RA) 19$^{h}$07$^{m}$49.60$^{s}$ and declination (dec) +62$^{\circ}$57$'$49.52$''$. Figure~\ref{HostGalaxy_Image} shows a pre-explosion image\footnote{Retrieved with \url{https://yymao.github.io/decals-image-list-tool/}} as well as an image of the supernova taken on the rise. 

A large amount of photometric and spectroscopic data was collected for SN\,2020qlb. This includes 11 spectra and 563 photometric observations, in 14 different filters, with coverage even during the SN's solar conjunction. In this section, we describe the data collection and reduction. The explosion date is estimated in Sect.~\ref{Explosion} to be MJD\,$59050.69 \pm{0.28}$.

\subsection{Photometry}

\subsubsection{ZTF photometry}\label{ZTFphotmetry}

The ZTF survey camera \citep{Dekany2020} is mounted on the Palomar Observatory Schmidt 48 inch Samuel Oschin telescope which scans the night sky in search of SNe and other interesting transients. On clear nights since 2018 it has been scanning more than 2750 square degrees per hour down to 20.5 mag in the $g$ and $r$ filters, and less frequently in $i$. Data-processing pipelines, alert systems and data archival, access and analysis are performed at  Infrared Processing and Analysis Center (IPAC) \citep{MasciEtal2019}, including image subtraction using the algorithm of \citet{ZackayEtal2016}. ``Forced'' point spread function (PSF) photometry is used to gather SN flux measurements in archived ZTF images, even from epochs prior to the transient's initial detection. \citet{YaoEtal2019} describe the ``forced photometry'' method and show that it is able to recover detections missed by the real-time pipeline as well as provide deeper predetection upper limits.

We performed the ZTF Forced PSF-fit Photometry service data reduction procedure (ver. 2.2)\footnote{\url{http://web.ipac.caltech.edu/staff/fmasci/ztf/forcedphot.pdf}} for each of the ZTF filter data sets wherein the baseline correction, the photometric uncertainty validation and the differential-photometry light curve generation were done accordingly to create observer frame light curves. There were 33 ZTF$_{g}$, 34 ZTF$_{r}$ but zero ZTF$_{i}$ pre-SN baseline values identified in the data set for SN\,2020qlb. We therefore estimated the baseline for the ZTF$_{i}$ filter so that its resulting apparent magnitudes agreed with the LT-$i$ filter measurements taken at overlapping phases. Finally, we excluded the 23 ZTF $g$-band measurements where the subtracted reference values were constructed from measurements taken during the supernova rise. All photometry is listed in Table~\ref{table:Photometry}.

We computed the absolute magnitude as
\begin{equation}\label{eq:7}
    \text{M} = \text{m} - \mu - K_{\text{corr}} - \text{A}_{\text{MW}} - \text{A}_{\text{host}}
\end{equation}
where $\mu$ is the distance modulus ($=39.40$ for z = 0.1583), $K_{\text{corr}}$ is the K-correction between the filter bandpass in the observer frame and the filter bandpass in the rest frame, A$_{\text{MW}}$ is the extinction from the Milky Way and A$_{\text{host}}$ is the extinction from the host galaxy.

The K-correction as described by \citet{HoggEtal2002} has two contributions: the first corrects for the redshift as
 \begin{equation}\label{eq:8}
    K_{corr} = -2.5 \times \log_{10}(1+z) = -0.16 \quad \text{mag},
\end{equation}
and the second corrects for the overall shape of the spectrum. We use only this first term of the K-correction; in practice, this means that all absolute magnitudes reported are at a bluer effective wavelength by a factor of ($1+z$) compared to the rest wavelength of the filter. We check the impact of this, for example by comparing peak absolute magnitudes by explicitly calculating the full K-correction from a spectrum taken near peak light using the {\tt SNAKE} code \citep{InserraEtal2018}. The result in $g$-band is still $K_g = -0.16 \pm 0.01~{\rm mag}$, so the peak $g$-band magnitude reported can be considered rest-frame.

We used the \citet{Fitzpatrick1999} extinction model to correct for the MW dust extinction based on the parameters R$_{V}=3.1$ and E(B-V)~$=0.053$ mag. In principle, we should also consider the possible extinction from the host galaxy. In the case of SN\,2020qlb, the host is a faint, blue dwarf galaxy, similar to typical SLSN-I host galaxies (Fig.~\ref{HostGalaxy_Image}; Sec.~\ref{sec:host}; \citealt{LunnanEtal2014,PerleyEtal2016}). The Balmer line ratios in the host galaxy spectrum (Sec.~\ref{sec:host}) indicate some host galaxy extinction ($E(B-V)_{\rm host} = 0.10 \pm 0.05~{\rm mag}$). However, since we cannot know whether the extinction of this H~II region is typical of the supernova site, we conservatively assume zero host galaxy extinction for the majority of the calculations in this paper. Where relevant, we point out how the results would change if host galaxy extinction was included.

\subsubsection{Swift UVOT photometry} \label{SwiftData}
We used the UV/Optical Telescope (UVOT) \citep{RomingEtal2005} on the Neil Gehrels (\textit{Swift}) Observatory. Measurements from six different filters, ranging from the ultraviolet (UV) to visible wavelengths, were retrieved from the NASA Swift Data Archive\footnote{ \url{https://heasarc.gsfc.nasa.gov/cgi-bin/W3Browse/swift.pl}} and processed using UVOT data analysis software HEASoft version 6.19\footnote{ \url{https://heasarc.gsfc.nasa.gov/}}. Source counts were then extracted from the images using a radius of 3 arcseconds, while the background was estimated using a radius of 48 arcseconds around the SN position. We then used the Swift tool \texttt{UVOTSOURCE} to obtain the count rates from the images before converting them to magnitudes using the UVOT photometric zero points \citep{BreeveldEtal2011a} and the September 2020 calibration files.

During 22 epochs, ranging between 26 and 143 days post explosion, all six filters were used separately to measure SN\,2020qlb's apparent AB magnitudes and their standard deviation of measurement error. Additional data are available where fewer than six filters were successfully utilized. We put all of the UVOT data into the rest frame using the procedure outlined in Sect.~\ref{ZTFphotmetry}. The observed photometry is listed in Table~\ref{table:Photometry}, and we plot the resulting rest frame UVOT light curves in Fig.~\ref{fig:AbsMag}.

\subsubsection{Liverpool Telescope photometry} \label{LiverpoolData}
The Liverpool Telescope (LT) \citep{SteeleEtal2004} has five filters ($u, g, r, i$ and $z$) available for photometric measurements on the optical imager IO:O. Reduced data were provided using the standard IO:O pipeline.

\begin{figure*}
        \centering
        \includegraphics[width=\textwidth]{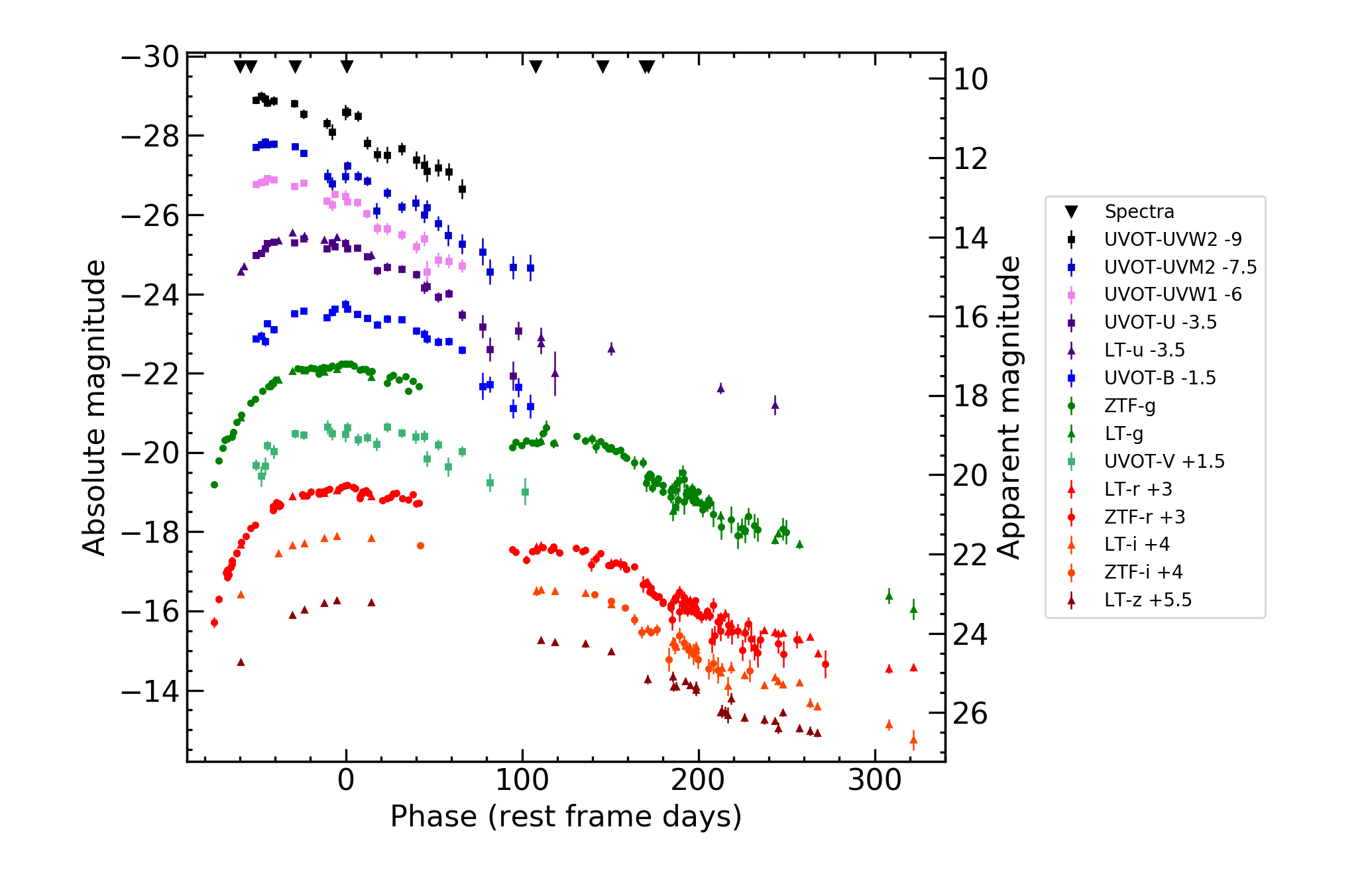}
        \caption{\label{fig:AbsMag}Absolute magnitude (rest frame) light curves from $\sim1800$ to 18500 $\AA$, including the Liverpool Telescope SDSS filters as triangles, the \textit{Swift} UVOT filters as squares and the ZTF filters as circles. The black triangles at the top indicate the phases of available spectra. Phase $=0$ (at peak ZTF $g$-band) is estimated in Sect.~\ref{GP}. The light curves in the different bands were shifted for illustration purposes.}
\end{figure*}

We observed apparent magnitudes between 17 and 398 days post explosion and put them into the rest frame using the procedure outlined in Sect.~\ref{ZTFphotmetry}. We list the resulting photometry in Table~\ref{table:Photometry} and plot the resulting rest frame LT light curves together with the UVOT and ZTF light curves in Fig.~\ref{fig:AbsMag}.

\begin{table}
\centering
\small
\caption{\label{table:Photometry} SN\,2020qlb photometric observations}
\begin{tabular}{ccccc}
\hline
\hline
MJD         & Phase$^{a}$        & Filter  & Brightness            & Telescope\\
            & (days)        &         &  (mag)         & +instrument\\
$($days)    & rest frame    &         &  rest frame            &  \\ \hline
59053.33     & -74.8        & r       &  20.66$\pm$0.14   &  P48+ZTF    \\
59053.37     & -74.8        & g       &  20.24$\pm$0.08   &  P48+ZTF    \\
59056.36     & -72.2        & r       &  20.06$\pm$0.10   &  P48+ZTF    \\
59056.41     & -72.1        & g       &  19.64$\pm$0.07   &  P48+ZTF    \\
59059.31     & -69.6        & g       &  19.32$\pm$0.07   &  P48+ZTF    \\
59060.36     & -68.7        & g       &  19.12$\pm$0.05   &  P48+ZTF    \\
59061.30     & -67.9        & r       &  19.39$\pm$0.07   &  P48+ZTF    \\
59062.27     & -67.1        & r       &  19.51$\pm$0.10   &  P48+ZTF    \\\hline
\end{tabular}

\tablefoot{The photometry is not corrected for reddening.
This table is available in its entirety in machine readable form and is also available on WISEREP}.\\
\tablefoottext{a}{relative to g-band maximum (MJD $59140.0$; see Section \ref{GP})}
\end{table}

\subsection{Spectra}
We acquired spectra with the SPectrograph for the Rapid Acquisition of Transients (SPRAT; \citealt{Piascik2014}) on the 2 m LT via the Transient Name Server (TNS)\footnote{\url{https://www.wis-tns.org/object/2020qlb}} \citep{TNS2020qlb}, the Spectral Energy Distribution Machine (SEDM; \citealt{BlagorodnovaEtal2018}) on the Palomar 60 inch telescope (P60), the Double Beam Spectograph (DBSP; \citealt{OkeGunn1982}) on the 200 inch Hale telescope (P200) at Palomar Observatory, the Low Resolution Imaging Spectrometer (LRIS; \citealt{OkeEtal1995}) on the 10 m Keck I telescope and the Andalucia Faint Object Spectograph and Camera (ALFOSC)\footnote{\url{http://www.not.iac.es/instruments/alfosc/}} on the 2.56~m Nordic Optical Telescope (NOT). Table~\ref{table:Spectra} lists available details about each of the spectra taken of SN\,2020qlb.

\begin{table}
\begin{center}
\small
\caption{\label{table:Spectra} Summary of SN\,2020qlb spectroscopic observations}
\begin{tabular}{lrlr}
\hline
\hline
MJD               & Phase$^{a}$ (days) & Telescope +        & Exposure\\
$($days)          & rest frame   &  instrument    & time (s)\\ \hline
59071.30          & -60          & P60 + SEDM               &  2250\\
59071.97          & -60          & LT + SPRAT $^{b}$      &   900\\
59075.38          & -57          & P60 + SEDM               &  2250\\
59078.40          & -54          & P60 + SEDM               &  2250\\
59107.20          & -29          & P200 + DBSP $^{d}$     &   900\\
59141.32          & +2           & KECK I + LRIS $^{c}$   &   300\\
59265.22          & +109         & NOT + ALFOSC $^{e}$    &  1350\\
59309.13          & +146         & NOT + ALFOSC $^{e}$    &  1350\\
59347.41          & +179         & P200 + DBSP $^{d}$     &  1200\\
59349.14          & +180         & NOT + ALFOSC $^{e}$    &  4500\\
59673.58          & +461         & KECK I + LRIS $^{c}$   &  2700\\\hline
\end{tabular}
\tablefoot{
\tablefoottext{a}{relative to g-band maximum (MJD $59140.0$; see Section \ref{GP})}
\tablefoottext{b}{Wasatch VPH model WP-600/600-25.4}
\tablefoottext{c}{blue grism 400/3400 and red grating 400/8500}
\tablefoottext{d}{blue 600/4000 and red 316/7500}
\tablefoottext{e}{Grism 4}
}
\end{center}
\end{table}

SPRAT spectra were acquired and automatically reduced according to Liverpool Observatory procedures\footnote{\url{https://telescope.livjm.ac.uk/TelInst/Inst/SPRAT/}}. SEDM spectra were reduced according to \citet{RigaultEtal2019}. DBSP spectra were reduced using a PyRAF-based pipeline described by \citet{BellmSesar2016}. The LRIS spectrum was reduced using LPipe as described by \citet{Perley2019}. ALFOSC spectra were reduced using PyNOT\footnote{\url{https://github.com/jkrogager/PyNOT}}. Each spectrum was calibrated against a spectrophotometric standard star. For the Keck spectrum at +461 days, we tied the flux scale to the host galaxy photometry. All spectra will be uploaded to WISEREP\footnote{\url{https://www.wiserep.org/}}.

\begin{figure*}
        \centering
        \includegraphics[width=\textwidth]{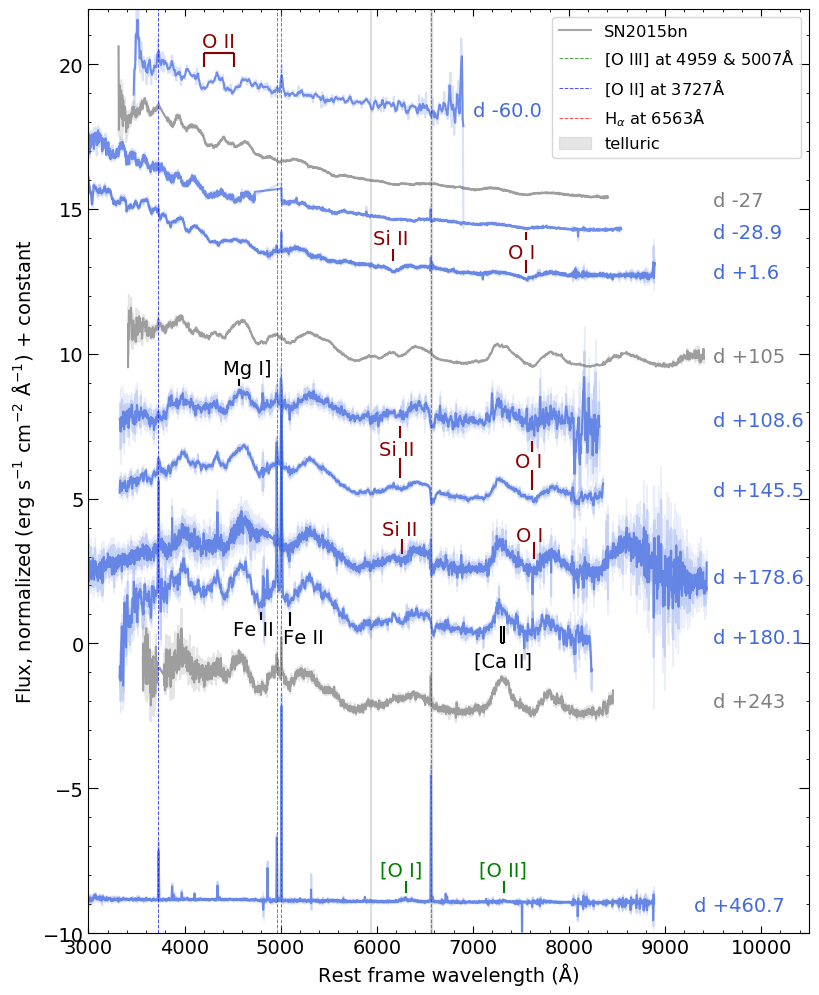}
        \caption{\label{SpectralEv}Spectral sequence of SN\,2020qlb. The dark blue SN\,2020qlb spectra were smoothed using a Savitzky-Golay low pass filter. A lighter shade of blue is used for un-smoothed data and measurement errors. Spectra from a well studied SLSN-I (SN\,2015bn) are shown in gray for comparison. P-Cygni profile minima used for velocity estimations are marked in dark red. Narrow host galaxy emission lines ([O\,III], H$\alpha$ and [O\,II]) used for redshift estimation are indicated with green, red and blue vertical dashed lines. Blended Fe II absorption features, as well as Mg I] and [Ca II] emission features are noted in black. Nebular phase [O\,I] and [O\,II] broad emission lines are noted in green.}
\end{figure*}

In Fig.~\ref{SpectralEv} we show sequential spectra from the different rest frame epochs together with relevant markings of important features. The noticeable gap between day +2 and day +109 includes a solar conjunction when ground based observations were not possible.

\subsection{X-ray observations}\label{XrayDetections}

While monitoring SN\,2020qlb with UVOT, \textit{Swift} also observed the field between 0.3 and 10 keV with its onboard X-ray telescope XRT in photon-counting mode \citep{Burrows2005a}. We analyzed these data with the online-tools of the UK \textit{Swift} team\footnote{\url{https://www.swift.ac.uk/user_objects/}} that use the methods described in \citet{Evans2007a, Evans2009a} and the software package \texttt{HEASoft} version 6.26.1.

SN\,2020qlb evaded detection in all epochs. The median $3\sigma$ count-rate limit of all epochs is 0.007~count~s$^{-1}$ (spread: 0.003 -- 0.03~count~s$^{-1}$) between 0.3--10~keV. Stacking all epochs pushes the $3\sigma$ count-rate limit to 0.0002~count~s$^{-1}$. To convert the count-rate limits into a flux, we assumed a power-law spectrum with a photon index $\Gamma$\footnote{The photon index $\Gamma$ is defined as the power-law index of the photon flux density ($N(E)\propto E^{-\Gamma}$).} of 2 and a Galactic neutral hydrogen column density of $5.75\times10^{20}$~cm$^{-2}$ \citep{HI4PI2016a}. Between 0.3--10 keV the median count-rate limits correspond to an unabsorbed flux of $2.9\times10^{-13}~{\rm erg\,cm}^{-2}\,{\rm s}^{-1}$ and luminosity of $2.0\times10^{43}~{\rm erg}\,{\rm s}^{-1}$ (if all observations are coadded); and $9.0\times10^{-15}~{\rm erg\,cm}^{-2}\,{\rm s}^{-1}$ and $6.2\times10^{41}~{\rm erg}\,{\rm s}^{-1}$ respectively (if dynamic rebinning is used).

\subsection{Host galaxy photometry} \label{HostGalPhotom}
We retrieved science-ready coadded images from the Sloan Digital Sky Survey data release 9 (SDSS DR 9; \citealt{Ahn2012a}), DESI Legacy Imaging Surveys \citep[Legacy Surveys, LS;][]{Dey2018a} data release 8, and the Panoramic Survey Telescope and Rapid Response System (Pan-STARRS, PS1) DR1 \citep{ChambersEtal2016}. We measured the brightness of the host using \program{LAMBDAR}\footnote{\href{https://github.com/AngusWright/LAMBDAR}{https://github.com/AngusWright/LAMBDAR}} \citep[\program{Lambda Adaptive Multi-Band Deblending Algorithm in R};][]{Wright2016a} and the methods described in \citet{Schulze2021a}. Table \ref{tab:hostphot} lists the measurements in the different bands.


\section{Light curve analysis}\label{light-curveProperties}

In this section we analyze the light curves. We estimate the explosion date, the epoch of the maximum $g$-band flux, characteristic light curve timescales, and the $g-r$ color evolution.

\subsection{Explosion date estimation} \label{Explosion}

To estimate the explosion date we first generated the flux light curves in Janskys from the observer frame arbitrary unit (\textit{DN}) flux and zeropoint (ZP) magnitudes for the ZTF forced photometry data.

We fit a Heaviside function multiplied by a power law, as done in \citet{MillerEtal2020}, to the complete set of baseline and early $g$- and $r$-band measurements to estimate the explosion date to be MJD\,$59048.2 \pm{1.8}$.

\begin{figure}
        \centering
        \includegraphics[width=\columnwidth]{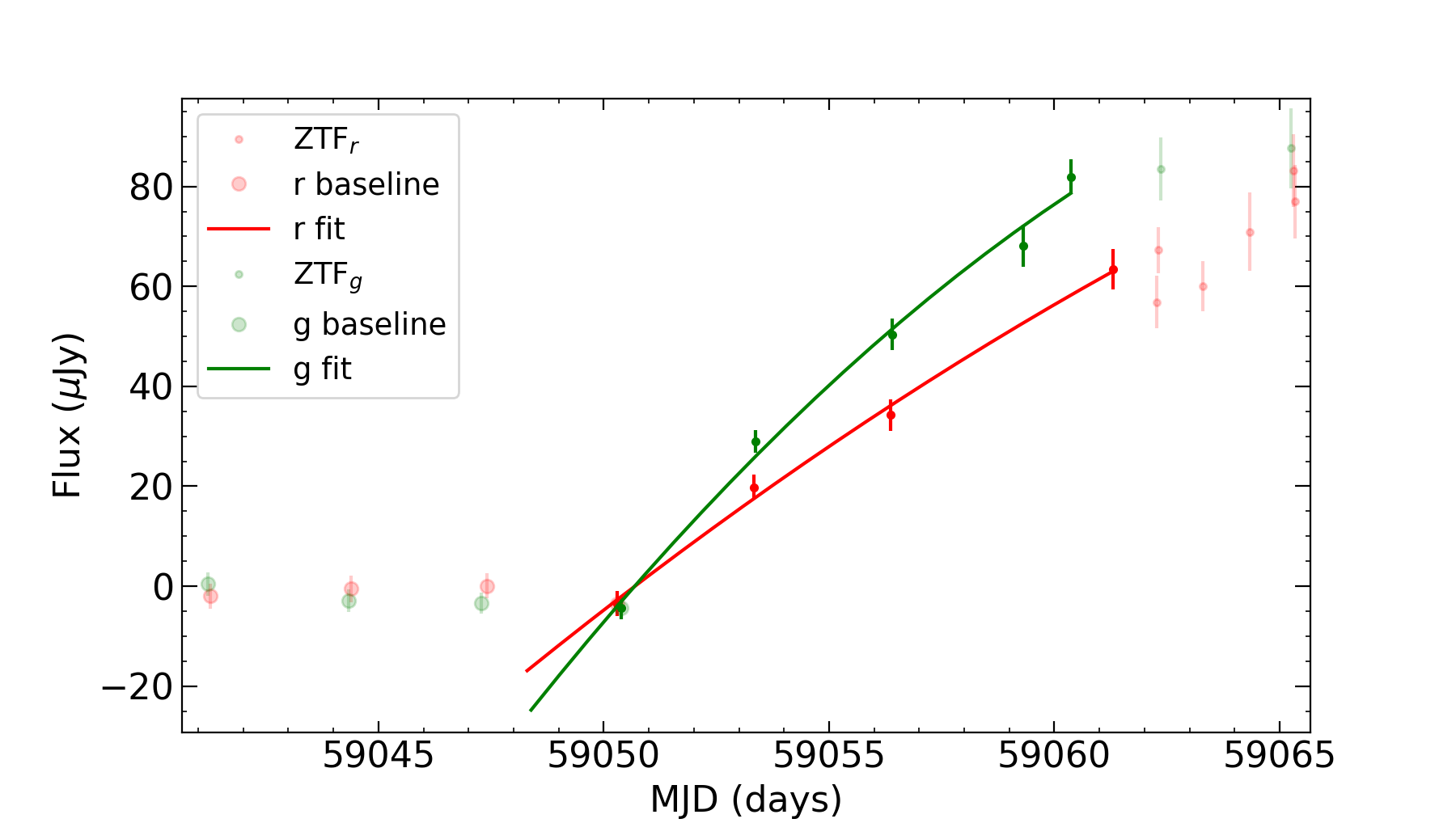}
        \caption{\label{fig:Expl} ZTF $r$- and $g$-band flux in the observer frame near in time to the initial detection of SN\,2020qlb. Points used for the second degree polynomial fits are darkened.}
\end{figure}

However, we found a better convergence by fitting (\texttt{numpy.polyfit}) a second order polynomial to both the first three $r$-band and to the first four $g$-band flux measurements that rise above the initial baseline values. Initially the $g$- and $r$-band fits both plotted earlier than their respective last baseline flux measurements, violating the upper limits these points provide. We therefore added the last baseline value for each filter band to the selected data points and refit the polynomials. The resulting fits are shown in Fig.~\ref{fig:Expl}.

We then ran a Monte-Carlo simulation of randomly selected data points from a Gaussian distribution of the one sigma uncertainties for each of the selected flux measurements. The resulting explosion date using the $g$ filter was MJD\,$59050.68\pm{0.20}$ and for the $r$ filter was MJD\,$59050.70 \pm{0.34}$. By combining both filter solutions we estimate an explosion date of MJD\,$59050.69 \pm{0.28}$.

The quoted error includes only the statistical uncertainty and not any systematic uncertainty terms. In practice, we have no constraints on the light curve below the ZTF detection limit ($\sim 21.1~{\rm mag}$ for ZTF $g$-band, corresponding to an absolute magnitude of $g ~\sim -18.3~{\rm mag}$). If SN\,2020qlb had an initial bump, plateau phase or a different rising slope below this limit it would not be captured by our uncertainty estimate for the explosion date.

\subsection{Peak g-band magnitude and light curve timescales} \label{GP}

In order to estimate the peak $g$-band magnitude and corresponding phase, as well as different measures of the rise and decline time, we interpolate the light curve. Following \citet{AngusEtal2019}, we use Gaussian Process (GP) regression interpolation utilizing the Python package \texttt{GEORGE} \citep{AmbikasaranEtal2015} with a Matern 3/2 kernel.

\begin{figure}
        \centering
        \includegraphics[width=\columnwidth]{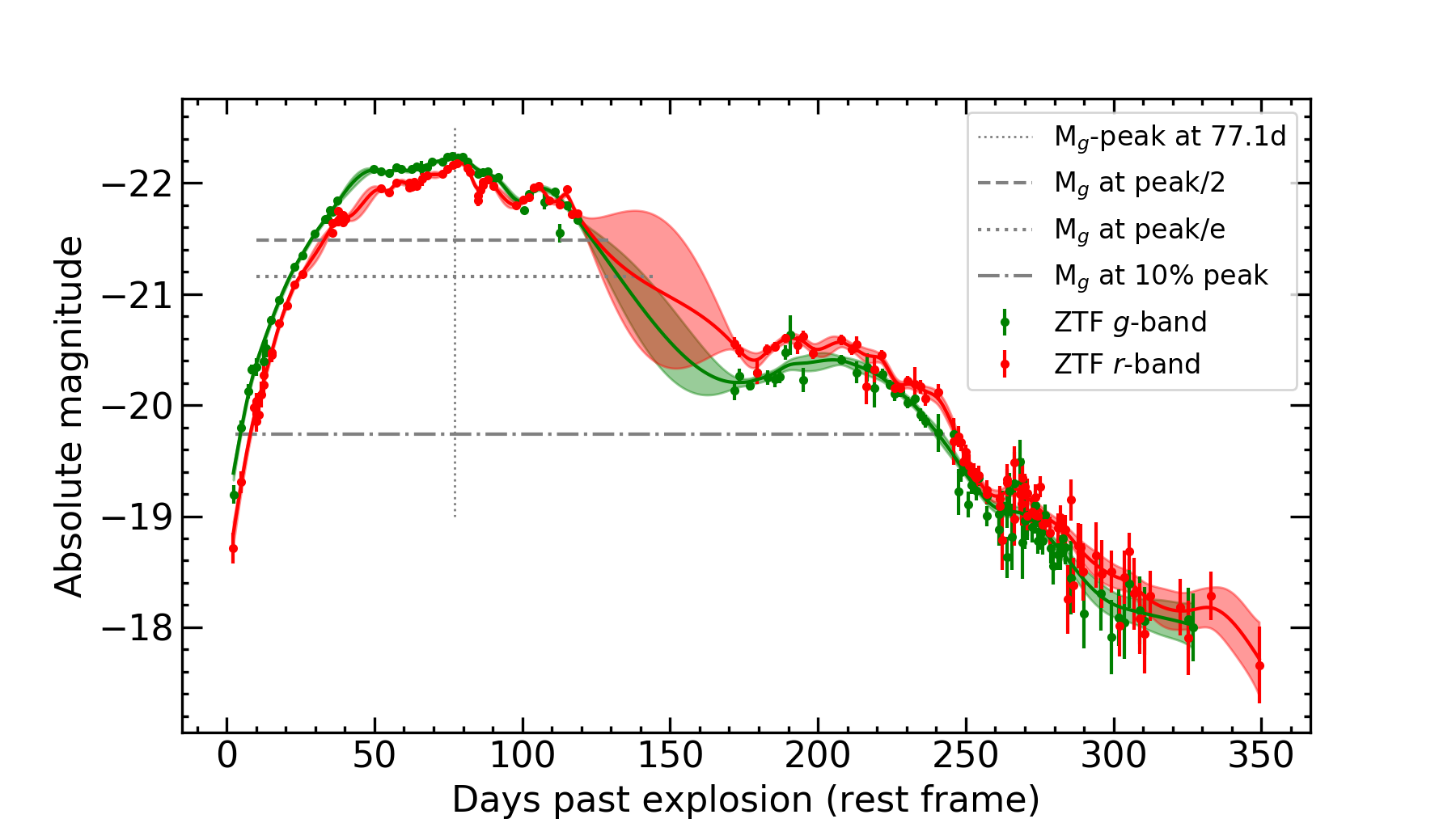}
        \caption{\label{fig:GPinterp}GP interpolated ZTF-$g$ and -$r$ (rest frame) light curves. Horizontal lines used to calculate the different rise and decline times are plotted in gray. The ballooning effect of the GP interpolation algorithm is clearly seen in the solar conjunction gap in the data between 120 and 170 days past explosion. }
\end{figure}

The resulting ZTF $g$- and $r$-band interpolated light curves are shown in Fig.~\ref{fig:GPinterp}. We used the interpolated curve to determine that the peak $g$ filter absolute magnitude occurred 77.1 days past explosion (MJD $59140.0$). The maximum M$_{g}$ was $-22.25\pm{0.01}$ mag. We note that this estimate does not include any correction for potential host galaxy extinction (Section~\ref{sec:EmissionLineDiag}); if this is also included the peak g-band magnitude would be $-22.62^{+0.11}_{-0.20}$ mag.

We then used the gray horizontal lines in Fig.~\ref{fig:GPinterp} together with the GP interpolated M$_{g}$ light curve to determine the rise- and decline times. The rest frame rise time from half maximum was $48.6\pm{0.5}$ days, from 1/e maximum was $55.7\pm{0.4}$ days, and from 10\% maximum was $72.4\pm0.3$ days. The rest frame decline time to half maximum was $46.6^{+4.6}_{-1.7}$ days, to 1/e maximum was $55.7^{+8.9}_{-4.5}$ days, and to 10\% maximum was $164\pm{2}$ days.

\begin{figure}
        \centering
        \includegraphics[width=\columnwidth]{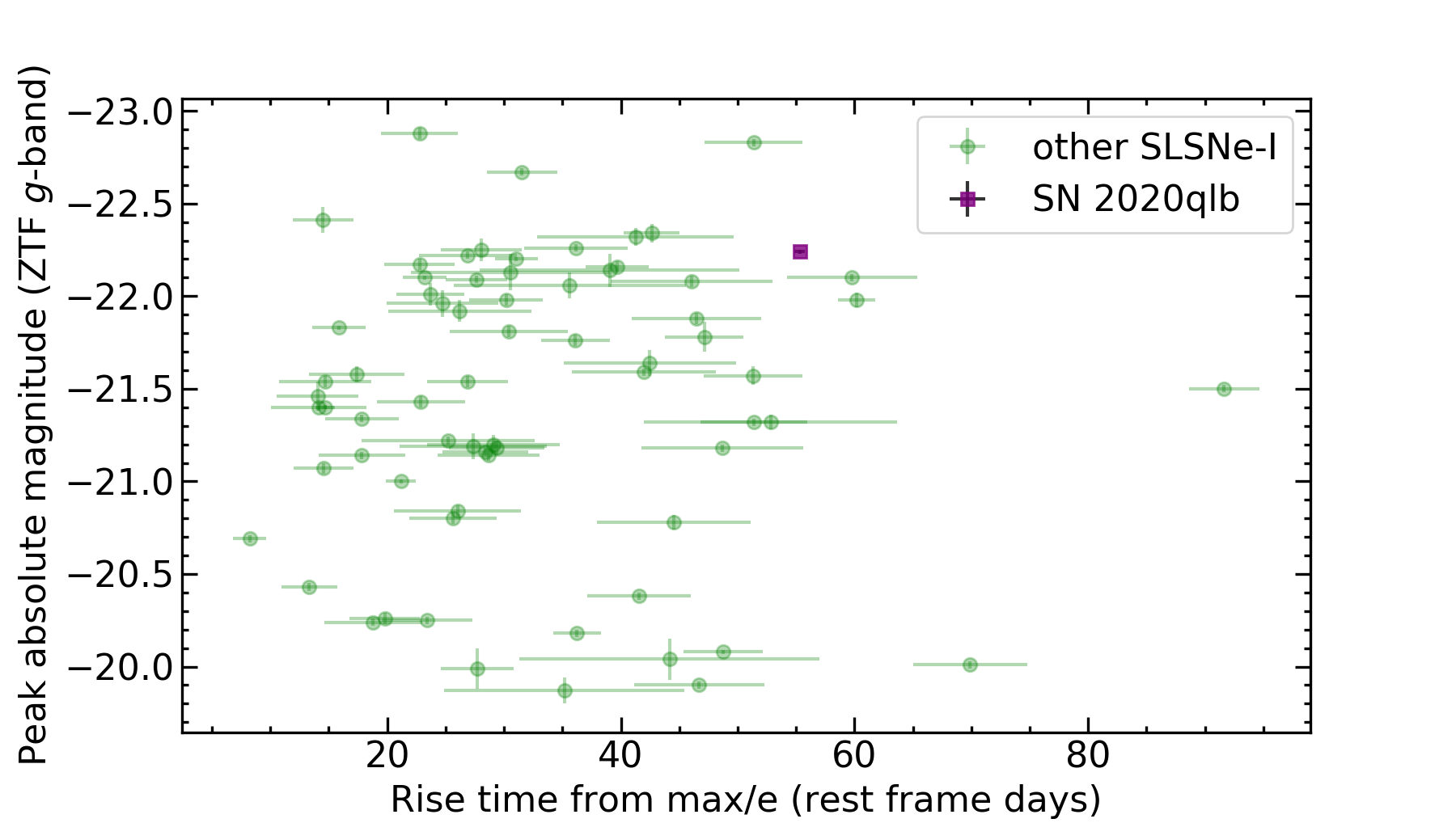}
        \caption{\label{MagGvsRiseTime}Peak M$_{g}$ versus e-folding rise time for 69 ZTF SLSNe-I from \citet{ChenEtal2022a} are compared with SN\,2020qlb.}
\end{figure}

To determine where SN\,2020qlb can be found in the phase space of rise time versus maximum absolute magnitude we plot the 69 SLSN-I from \citet{ChenEtal2022a} in comparison with SN\,2020qlb in Fig.~\ref{MagGvsRiseTime}. The rise time and peak brightness of SN\,2020qlb are both high (94th and 89th percentiles respectively) among SLSNe-I, but not unprecedented.

\subsection{g-r color evolution} \label{ColorEvolution}
\begin{figure}
        \centering
        \includegraphics[width=\columnwidth]{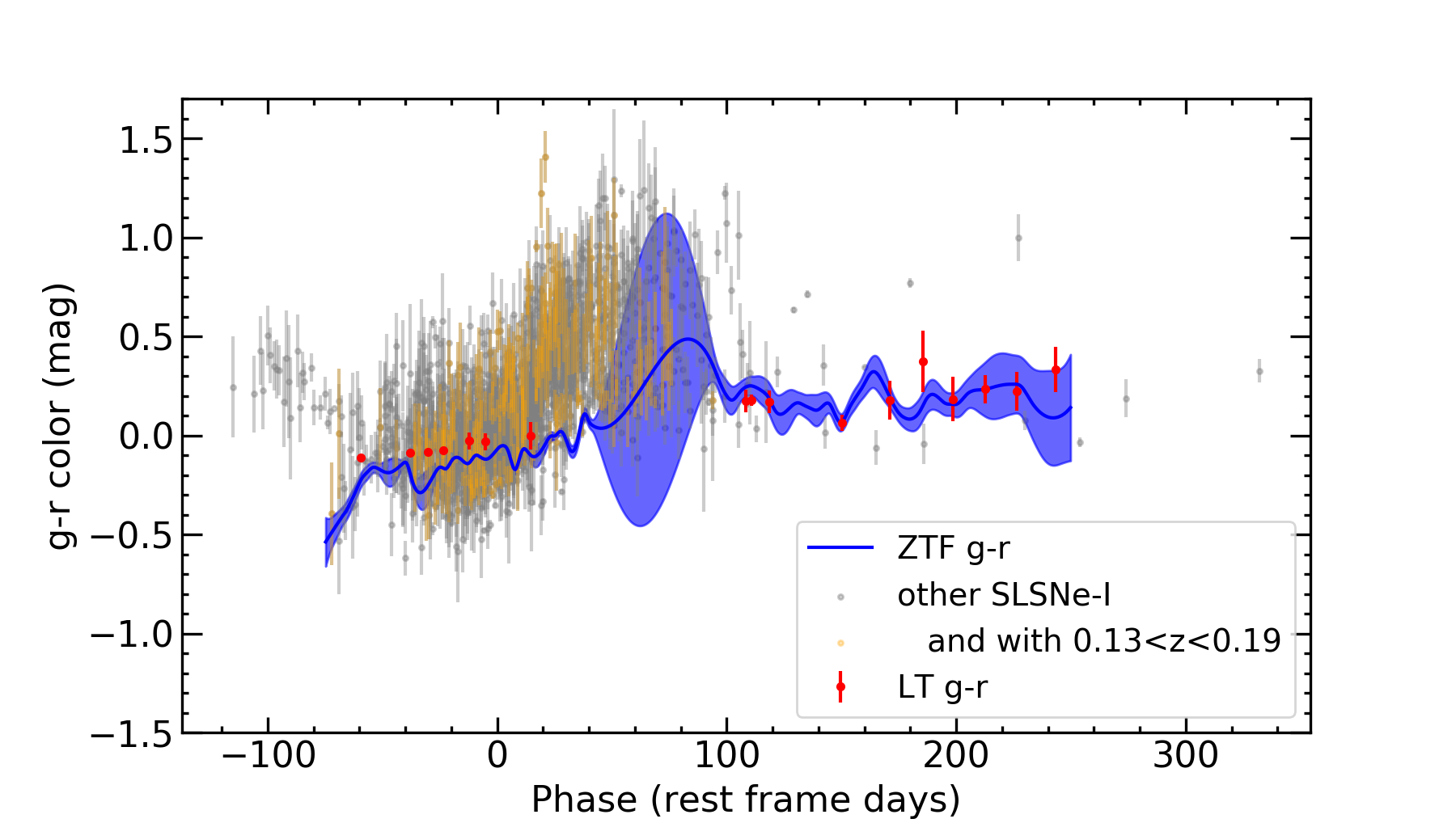}
        \caption{\label{fig:grEvol}SN\,2020qlb $g-r$ color evolution created from the ZTF-$g$ and -$r$ filter GP interpolations in Sect.~\ref{GP} in blue and the g-r measurements with the LT in red. The gray background is a scatter plot of the 75 SLSNe-I and orange is for the 9 with $0.13 <$ z $< 0.19$ from \citet[Fig. 13]{ChenEtal2022a}.}
\end{figure}

We also use the GP interpolated ZTF $g$- and $r$-band light curves to construct the $g-r$ (color) magnitude evolution plot shown in comparison to a recent survey of SLSNe-I \citep{ChenEtal2022a} in Fig.~\ref{fig:grEvol}. SN\,2020qlb's color evolution is one of the bluest but otherwise evolves normally compared to other SLSNe-I. We note that this color curve is calculated assuming zero host galaxy extinction; if this is included (Section ~\ref{sec:EmissionLineDiag} it would shift the color of SN\,2020qlb even bluer by another 0.13~mag.

\section{Spectral analysis} \label{SpectralProperties}

In this section we analyze SN\,2020qlb's spectral evolution shown in Fig.~\ref{SpectralEv} to estimate the host galaxy redshift, the SN spectral classification and the ejecta velocity evolution. We also note that there are no typical narrow lines which are signature spectral CSM interaction features (see Sect.~\ref{CSMinteraction}) at any phase.

\subsection{Host galaxy redshift} \label{HostGalaxyRedShift}
We determine the SN host galaxy redshift ($z$) by using the H$\alpha$ line in the rest frame -28.5d spectrum, the galaxy's strong narrow forbidden [O\,III] transitions at 4959Å and 5007Å as well as the [O\,II] transition at 3727Å. We thereby infer a redshift of $z$~=~0.1583.

\subsection{Spectral classification} \label{SpectralClassification}
We confirm the SLSN-I classification of SN\,2020qlb by comparing the hot and cool photospheric phase \citep{Gal-Yam2019} spectra to well studied SLSN-I spectra.

In the upper three spectra in Fig.~\ref{SpectralEv} we find the typical O\,II ``W'' feature around 4500Å as well as a characteristic blue continuum. The three spectra from SN\,2020qlb compare well with the hot photospheric phase spectrum from SLSN-I 2015bn at phase -27 days \citep{NichollEtal2016}.

We also compare the lower four spectra from SN\,2020qlb in Fig.~\ref{SpectralEv} with two cool photospheric phase spectra from SLSN-I 2015bn. Prevalent matching features such as the Mg I] and [Ca II] broad emission lines, noted in Fig.~\ref{SpectralEv}, indicate that SN\,2020qlb evolved as a typical SLSN-I in the cool photospheric phase between phases +109 and +180.

\subsection{P-Cygni velocity estimations} \label{PCygniVelocities}
P-Cygni profiles were found in the SN\,2020qlb photospheric phase spectra for the Si~II, O~II and O~I lines marked in dark red in Fig.~\ref{SpectralEv}. We used these profiles to estimate the ejecta velocity according the longitudinal relativistic Doppler shift. We fit an equation consisting of a straight line component to match the continuum, and a Gaussian component to match the absorption feature to P-Cygni profile data using the \texttt{scipy.optimize.curve\_fit} algorithm. The resulting $\lambda_{min}$ parameter fit values and covariance matrices were then used to produce absorption line velocity estimates and their propagated errors. The resulting estimates are plotted in Fig.~\ref{VelocityEv}.

We find that the maximum velocity is $\sim10000$ km~s$^{-1}$, the velocity near peak is $\sim8000-10000$ km~s$^{-1}$ and $\sim4000-600$0 km~s$^{-1}$ at $\gtrsim100$ days post-peak. \citet[][Fig.1]{ChenEtal2022b}, in 56 events, find that SLSNe-I have typical near peak O~II velocities of $\sim12000$ km~s$^{-1}$, ranging between $\sim6000$ and $\sim21000$ km~s$^{-1}$.

\section{Photospheric temperature and radius} \label{Blackbodyfits}

To measure the photospheric temperature and velocity evolutions we GP interpolate all light curves and extract the values at the time of the UVOT observations, and fit each epoch with a Planck function. We then utilize the \texttt{scipy.optimize.minimize}\footnote{\url{https://docs.scipy.org/doc/scipy/reference/generated/scipy.optimize.minimize.html}} algorithm to fit a blackbody to the data at each of the 22 UVOT epochs, as well as the 17 full sets of LT data available at later epochs. The algorithm therein estimates the best-fit temperature and radius at each epoch including a one standard deviation error for both.

\begin{figure}
        \centering
        \includegraphics[width=\columnwidth]{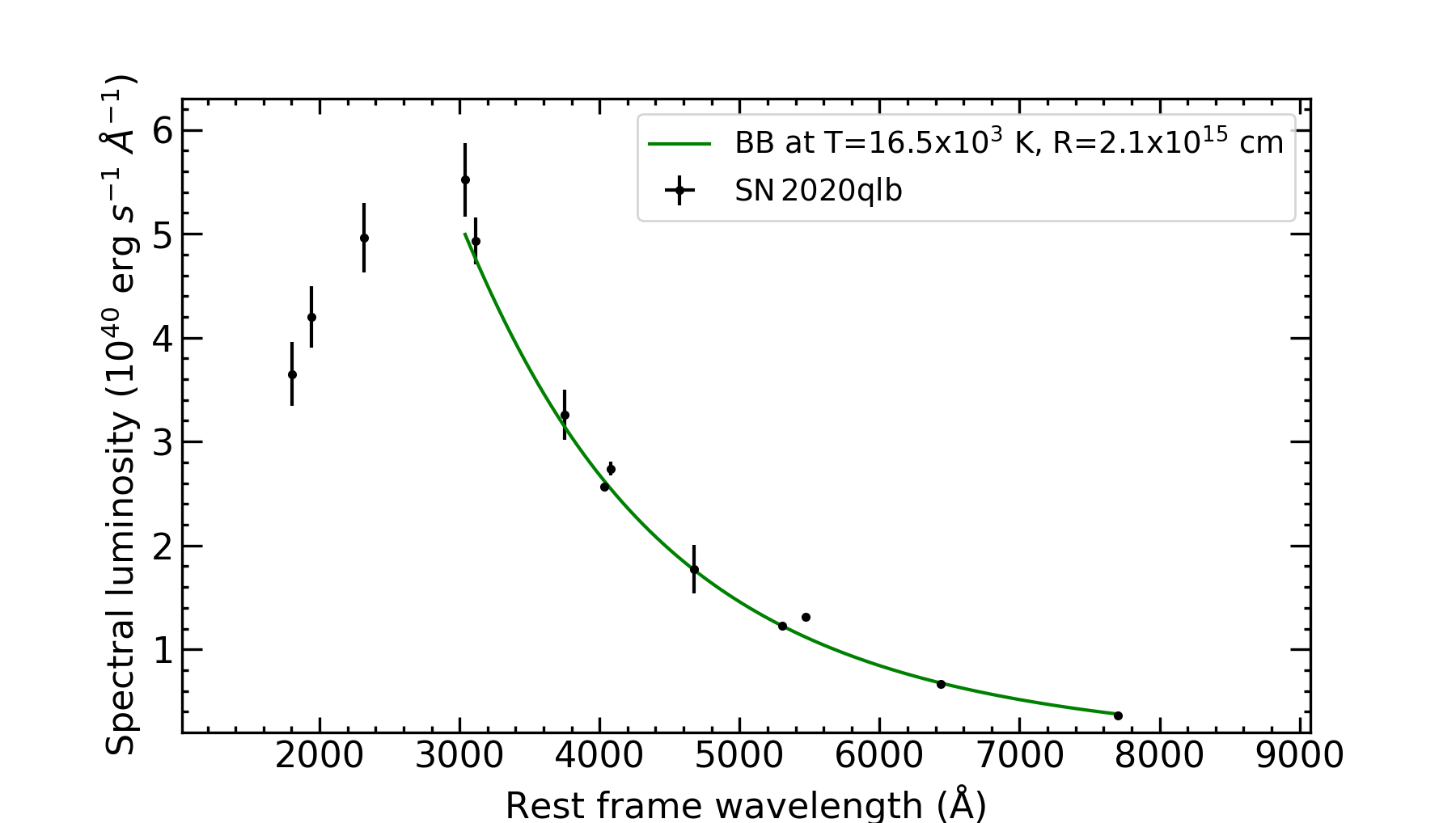}
        \caption{\label{4fits}Example (phase -51.2 days) of how a blackbody fit to photometric data is used to estimate the temperature and photospheric radius. The three UVOT UV filter measurements are excluded due to the UV line-blanketing effect.}
\end{figure}

The optical data are adequately described by the Rayleigh-Jeans tail of the Planck function. However, the three UV filters are consistently incompatible with the best-fit blackbody function, which we attribute to line blanketing. We note that this effect is not uncommon in SLSNe \citep[see e.g.,][]{YanEtal2017b}; a similar choice is done by \citet{Nicholl2018} in their \texttt{SuperBol} software. We therefore exclude these three filters when fitting the blackbody temperatures and radii. Fig.~\ref{4fits} shows an example of the resulting blackbody fit to the selected data at phase -51.2 days relative to the M$_{g}$ maximum.

\begin{figure}
        \centering
        \includegraphics[width=\columnwidth]{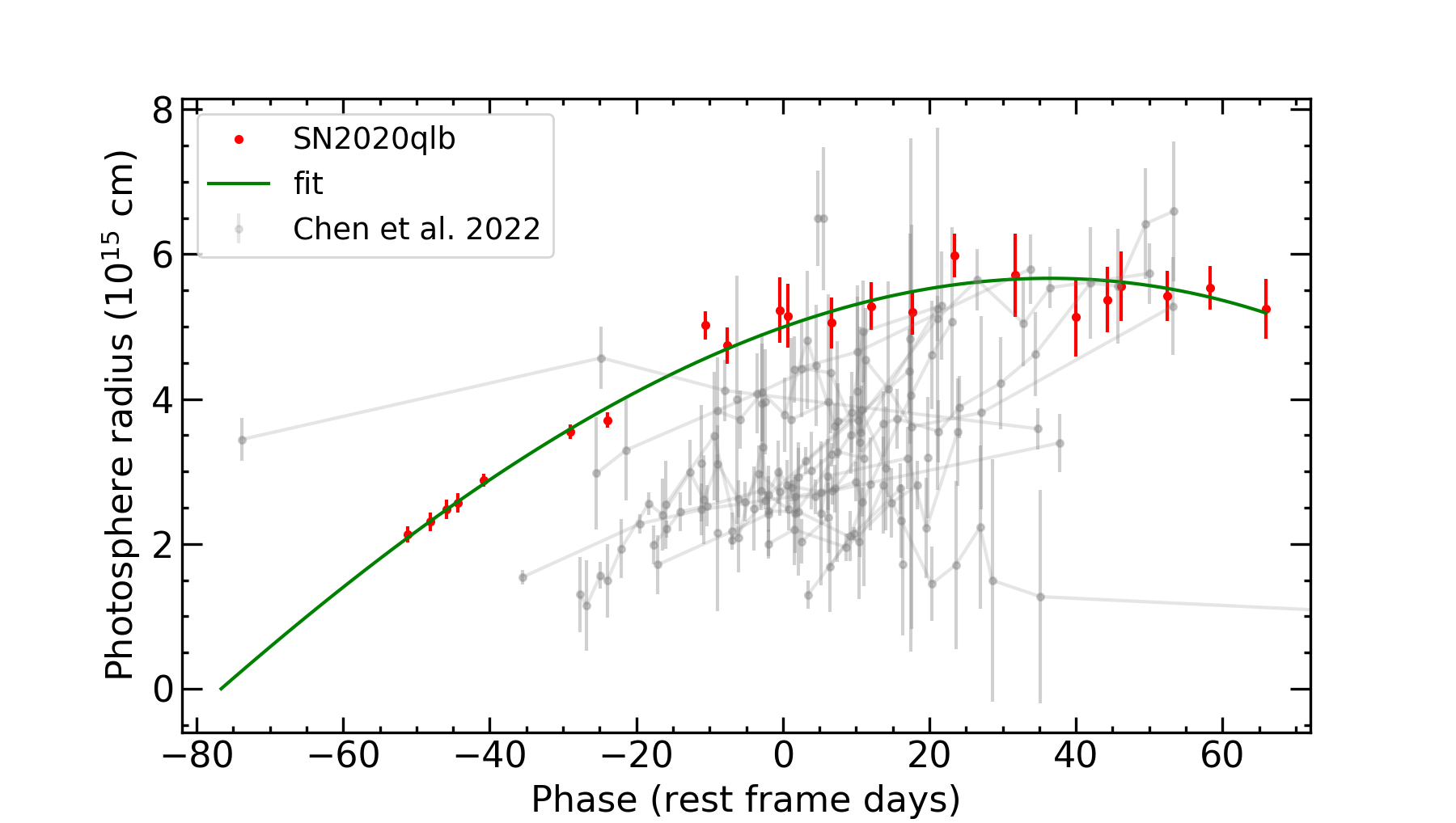}
        \caption{\label{RadiusEv}Rest frame radius evolution of the SN photosphere from blackbody fits to data from UVOT epochs. A fourth degree polynomial was fit to the data. SLSN-I radius evolutions from \citet{ChenEtal2022a} are shown in gray.}
\end{figure}

The resulting evolution of SN\,2020qlb's photospheric radius is plotted in Fig.~\ref{RadiusEv}. It is generally comparable to the 31 SLSNe-I from \citet{ChenEtal2022a}.

\begin{figure}
        \centering
        \includegraphics[width=\columnwidth]{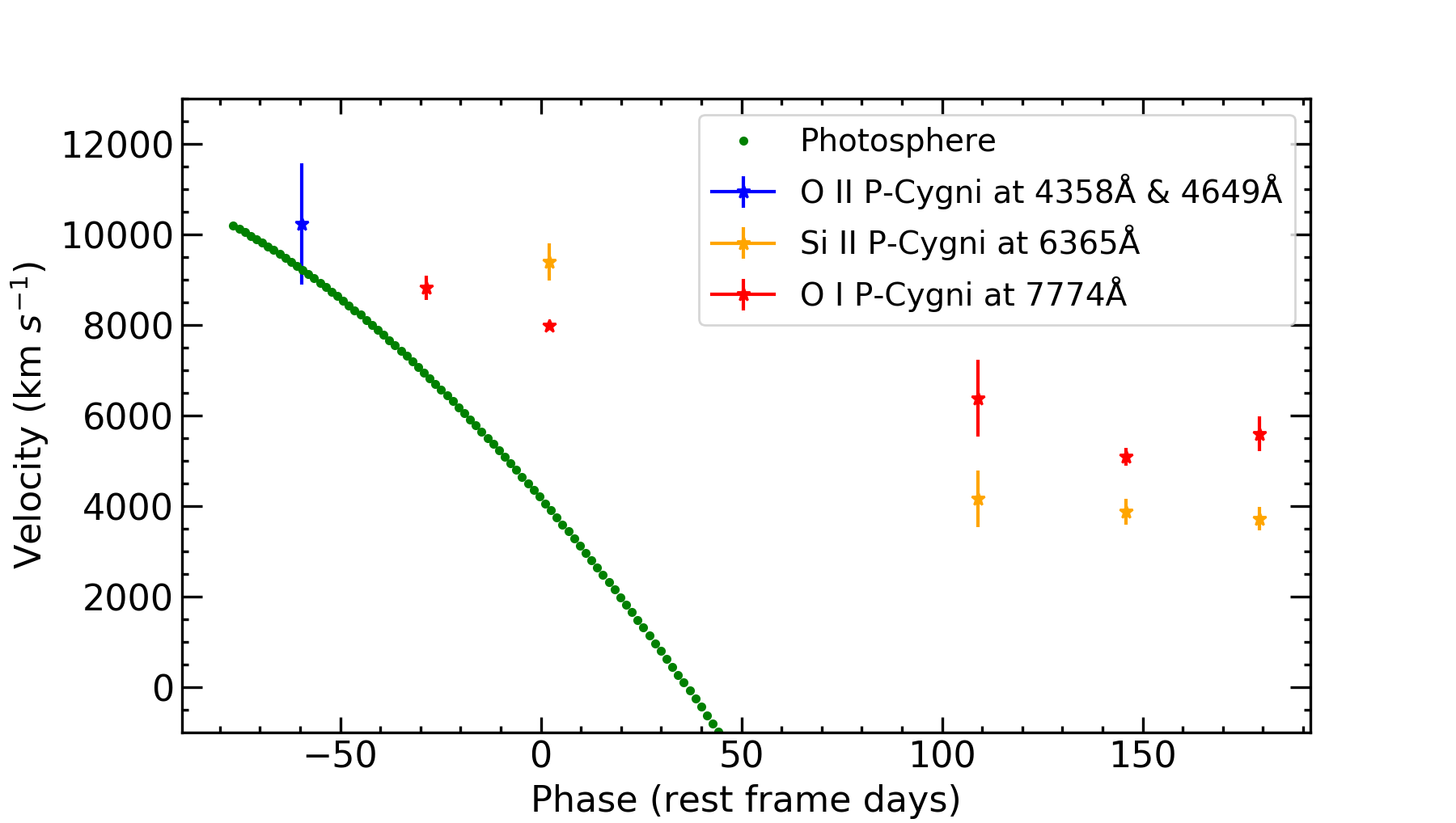}
        \caption{\label{VelocityEv}Velocity evolution of the SN photosphere shown in green based on the fourth degree polynomial fit to the photosphere radius evolution shown in Fig.~\ref{RadiusEv}. P-Cygni absorption line velocity estimates from Sect.~\ref{PCygniVelocities} are included as well.}
\end{figure}

We estimate the photosphere velocity evolution by plotting the derivative of the fourth degree polynomial fit of the radius evolution in Fig.~\ref{VelocityEv} as well as the velocity estimates from the P-Cygni profiles from Sect.~\ref{PCygniVelocities}. A general convergence of the maximum velocity estimates in Fig.~\ref{VelocityEv} appears to be $\sim10000$ km~s$^{-1}$.

\begin{figure}
        \centering
        \includegraphics[width=\columnwidth]{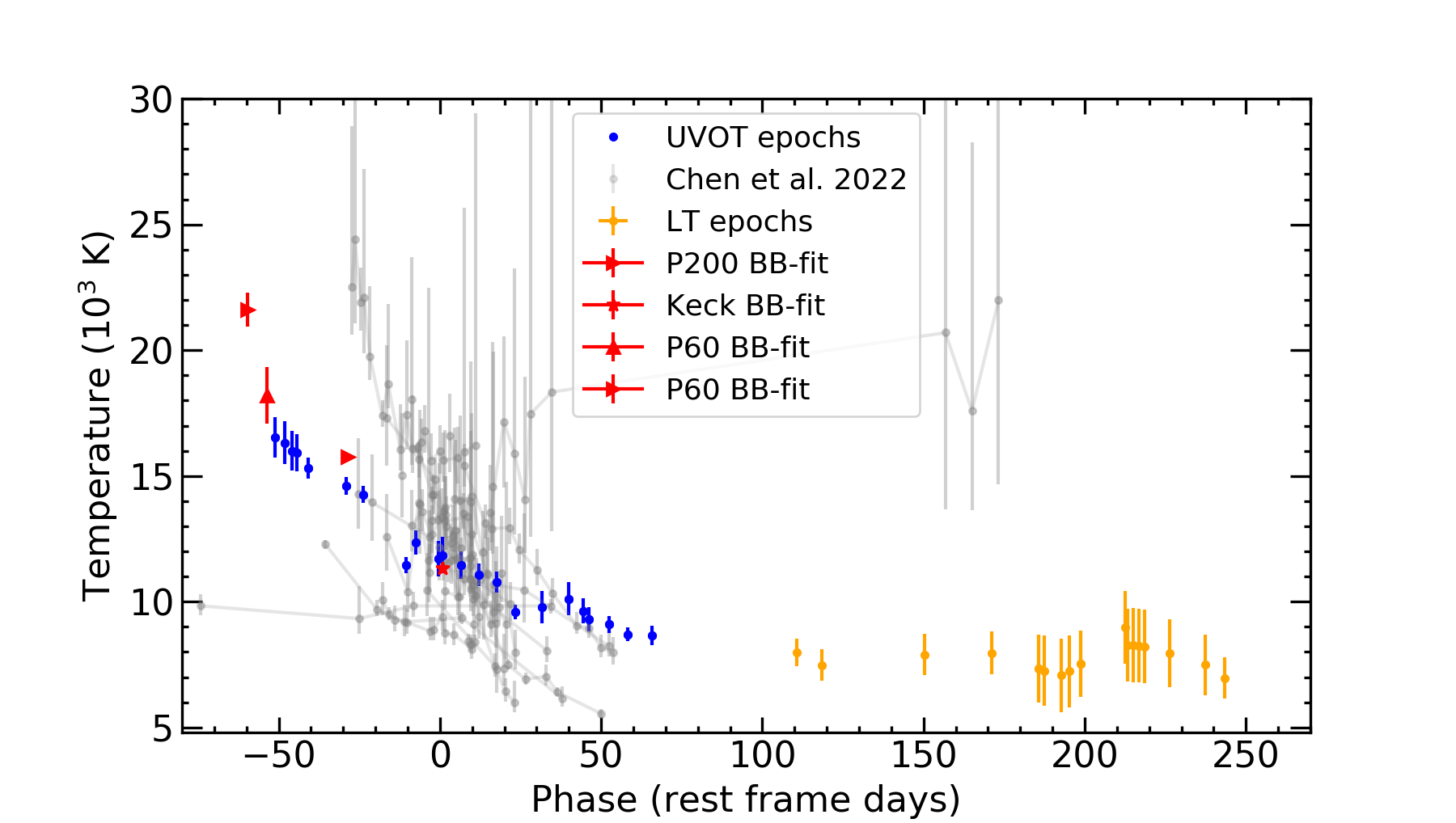}
        \caption{\label{TempEv}Temperature evolution of the SN\,2020qlb photosphere (blue: fit to UVOT + optical data; orange: fit to LT ugriz data; red: fit to spectra) in comparison to the 31 SLSN-I temperature evolutions from \citet{ChenEtal2022a} shown in light gray. }
\end{figure}

We also estimate SN\,2020qlb's temperature from spectra that were taken during the early photospheric phase of the SN. A blackbody was fit to the rest frame spectral data. The resulting temperature estimations are plotted in red in Fig.~\ref{TempEv} together with the other temperature estimates in blue and orange as well as with the 31 SLSNe-I from the ZTF's phase-I survey \citep{ChenEtal2022a} in light gray. SN\,2020qlb's temperature evolves similarly to most of the 31 SLSNe-I.

\section{Bolometric light curve} \label{Bololight-curve}

To construct a bolometric light curve we first estimate SN\,2020qlb's spectral luminosity ($L_{\lambda}$) at all wavelengths at each epoch. The spectral energy distributions (SEDs) at each epoch are then integrated over all wavelengths to create the bolometric light curve.

\subsection{Spectral energy distributions} \label{SEDs}

Given a set of rest frame $L_{\lambda}$s calculated from measurements in filters ranging from the UV to the visible we interpolate straight lines between data points. At wavelengths lower than the lowest filter wavelength and at wavelengths higher than the highest filter wavelength we use two similar sets of extrapolation methods.

\begin{figure}
        \centering
        \includegraphics[width=\columnwidth]{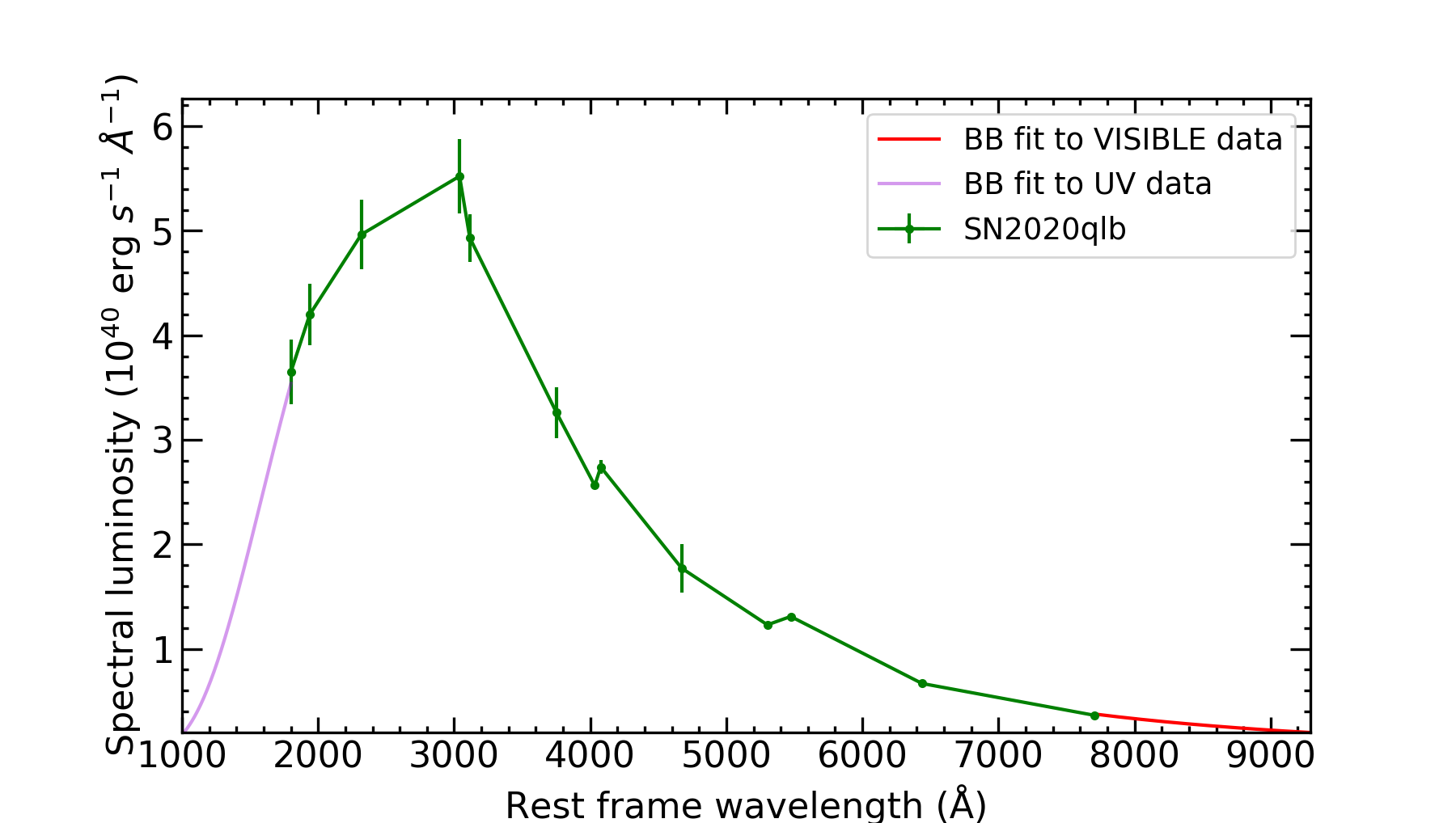}
        \caption{\label{SEDexample}Example, at a phase of -51.2 days, of how an SED is constructed using linear interpolations between photometric data points and extrapolated on both ends as explained in the text.}
\end{figure}

At the 22 epochs where UVOT filter data is available we create the extrapolation short of the shortest wavelength by fitting a blackbody to LT's $u$-band and the UVOT's UV filters, while a blackbody fit to the optical bands is used to create the extrapolation long of the longest wavelength (see also Sect.~\ref{Blackbodyfits}). These methods are also used in \texttt{SuperBol} software as described by \citet{Nicholl2018}. Fig.~\ref{SEDexample} shows an example of how the SED is constructed for phase -51.2 days from the M$_{g}$ peak.

At later epochs where a complete set of the five LT filter data is available we use the extrapolation methods described by \citet{LymanEtal2014} to create the SEDs. These extrapolation methods differ from the \texttt{SuperBol} methods in that we draw a straight line between LT's $u$-band and L$_{\lambda}$=0 at 2000Å on the side short of the shortest wavelengths. At additional epochs, where only 3 or 4 filter measurements are available, we use a GP interpolation (see Sect.~\ref{GP}) to estimate the missing filter magnitude(s) and error(s) before constructing the SEDs.

The maximum $L_{\lambda}$ is encapsulated by the data at each epoch; an example of this is shown in Fig.~\ref{SEDexample}. We therefore expect that a significant amount of the bolometric flux is included within the measurement ranges and that the two extrapolation methods to create the SEDs are accurately estimating the bolometric luminosities at their respective phases.

\subsection{Bolometric luminosities} \label{BoloLums}
\begin{figure}
        \centering
        \includegraphics[width=\columnwidth]{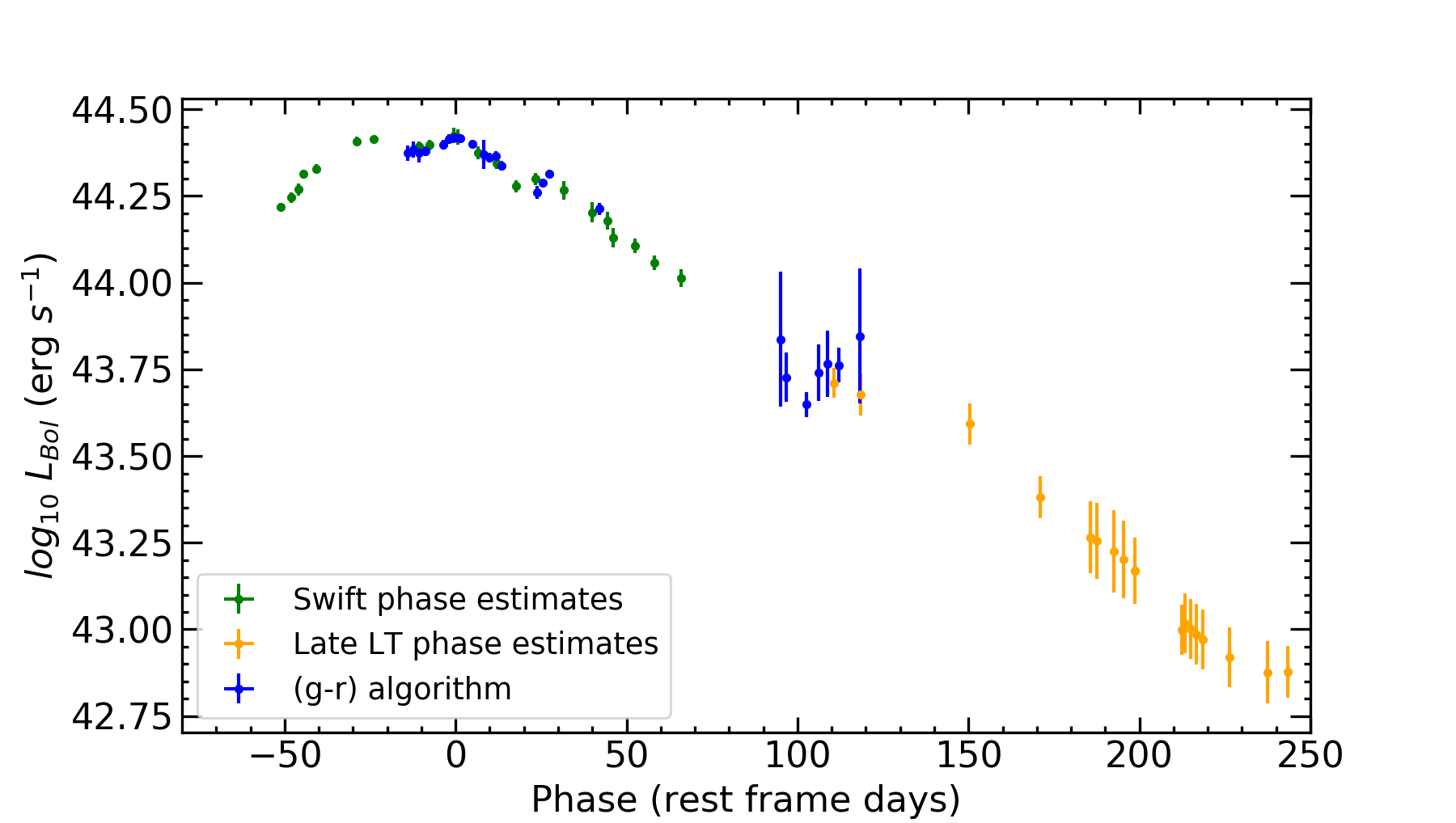}
        \caption{\label{BoloLCinterpolated}Interpolated bolometric luminosities are shown interlaced with the bolometric luminosities estimated in Sect.~\ref{BoloLums}.}
\end{figure}

At each of the 39 epochs where SEDs were created (see Sect.~\ref{SEDs}) we calculated the bolometric luminosity by integrating over wavelength. We then performed a Monte Carlo simulation, using samplings from a normal distribution of the $L_{\lambda}$ errors, to create error values for the luminosity estimates. The resulting plot of the bolometric luminosities and their errors over time are plotted in green (UVOT phases) and yellow (LT phases) in Fig.~\ref{BoloLCinterpolated}.

\subsection{Bolometric interpolations} \label{BoloInterpol}
The ZTF has a higher cadence of measurement than the \textit{Swift}/UVOT telescope. We therefore linearly interpolate luminosities at epochs between the bolometric data points created in Sect.~\ref{BoloLums}.

\citet{LymanEtal2014} created analytic relationships to estimate bolometric light curves for different types of core collapse SNe by identifying correlations between different measurements of color, for example M$_{g}$-M$_{r}$ $\equiv (g-r)$, and a bolometric correction factor to the absolute magnitude in the $g$-band BC$_{g}$ $\equiv$ M$_{\text{bol}}$ - M$_{g}$, where M$_{\text{bol}}$ is defined as follows:
\begin{equation}\label{eq:14}
    \text{M}_{\text{bol}} \equiv \text{M}_{\odot,\text{bol}} - 2.5 \times \text{log}_{10}\Bigg( \frac{\text{L}_{\text{bol}}}{\text{L}_{\odot,\text{bol}}} \Bigg)
\end{equation}
where L$_{\text{bol}}$ is the bolometric luminosity, L$_{\odot,\text{bol}}$ and M$_{\odot,\{bol}$ are the sun's luminosity (L$_{\odot,\text{bol}}=3.828\times10^{33}$ erg/s) and bolometric magnitude (M$_{\odot,\text{bol}}=4.74$ in the $g$-band) \citep{MamajekEtal2015}.

\begin{figure}
        \centering
        \includegraphics[width=\columnwidth]{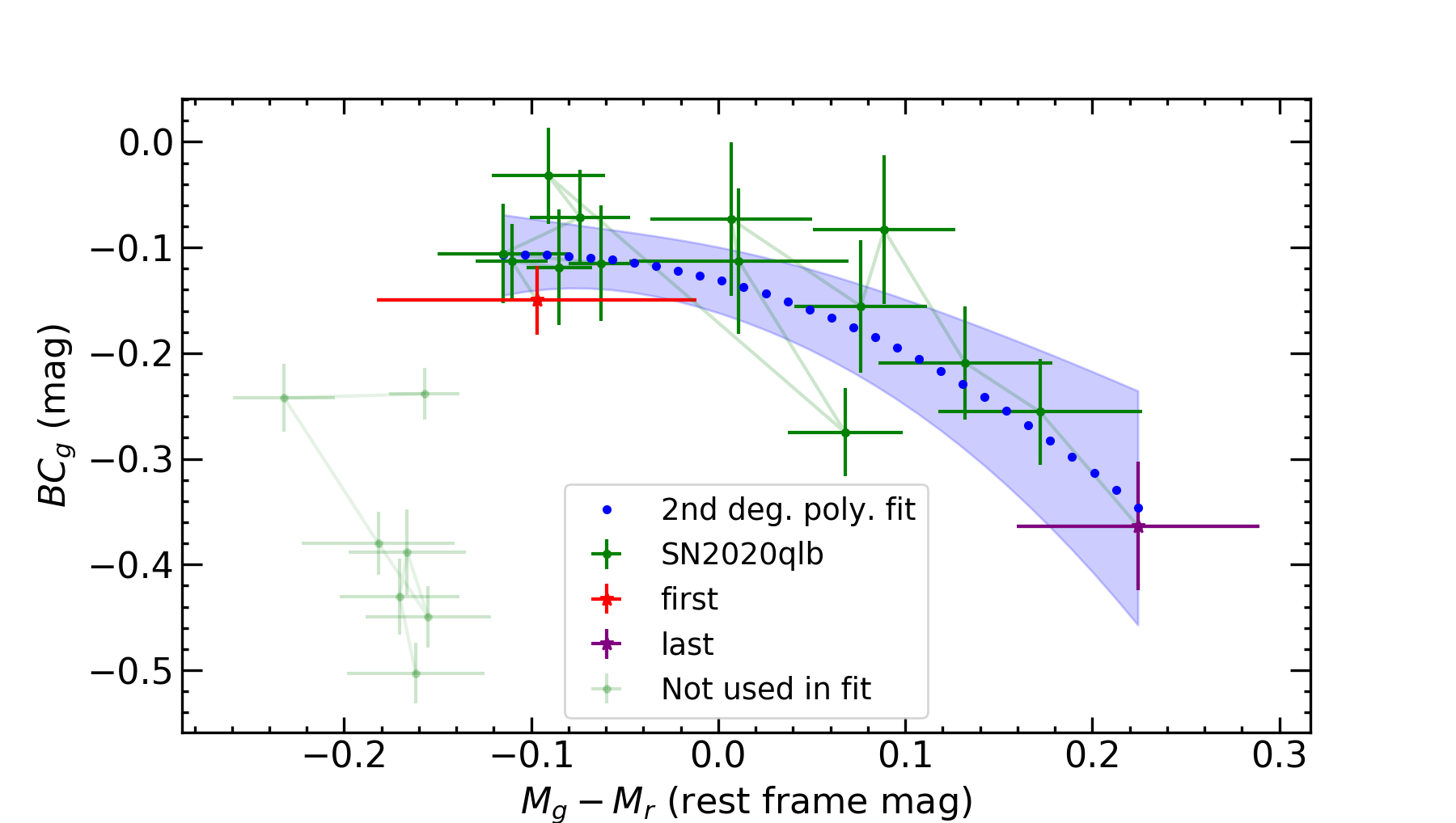}
        \caption{\label{BCg_vs_color_fit}Bolometric correction as a function of rest frame g-r color. A second degree polynomial curve-fit is compared to the post peak bolometric correction to the $g$-band ($BC_{g}$) versus $(g-r)$ color data. The data are shown connected in time sequence. The first seven points were removed from the curve-fit.}
\end{figure}

Using the \textit{Swift} and LT phase bolometric luminosities, interpolated ZTF-$g$ and -$r$ measurements as well as calculated absolute $g$-band magnitudes we plot BC$_{g}$ versus $(g-r)$ as shown in Fig.~\ref{BCg_vs_color_fit}. The first seven data points in Figure~\ref{BCg_vs_color_fit} show no relationship between the plotted parameters, likely because at these early epochs the SED peaks sufficiently far into the UV so that the $g-r$ color is not affected. We therefore exclude these epochs from our fit.
 A second degree polynomial using the \texttt{scipy.optimize.curve\_fit} algorithm produces the fit (in blue) shown in Fig.~\ref{BCg_vs_color_fit} for the post-peak data.

The resulting relationship between $BC_{g}$ and $(g-r)$ is:

\begin{eqnarray*}
    \text{BC}_{g} & = & -(2.340\pm1.419)\times(g-r)^2 - (0.494\pm0.142)\times\\
    & & (g-r) - (0.1314\pm0.023) \quad \text{mag}.
\end{eqnarray*}
We then used ZTF-g and -r measurements made after the phase of the last removed $(g-r)$ value and where $(g-r)<0.3$ to provide the interpolated bolometric luminosities plotted in blue in Fig.~\ref{BoloLCinterpolated}. 

By integrating the entire bolometric light curve in Fig.~\ref{BoloLCinterpolated} using a simple trapezoidal method we estimate the total radiated energy of SN\,2002qlb to have been $>2.8\pm{0.05}\times10^{51}$ erg. Integrating only over the observed filters (i.e., no extrapolations at either the red or the blue end, and not extrapolating to the presumed explosion date, see Sec.~\ref{SEDs}) gives a strict lower limit on the total radiated energy of $2.1 \times 10^{51}~{\rm erg}$. We note that this bolometric light curve integration is calculated assuming zero host galaxy extinction; if this is included (see Section ~\ref{sec:EmissionLineDiag}) the total radiated energy would be $>2.9\pm{0.1}\times10^{51}$ erg; and $>2.2 \times 10^{51}~{\rm erg}$ if no extrapolations are used.

\section{Light curve modeling} \label{Models}

In this section, we use the light curves (observed and bolometric) as well as the measured velocity to compare SN\,2020qlb to different semi-analytic models in order to determine the most likely power source.

\subsection{Radioactive source model (Arnett)} \label{RadioactiveModel}
\citet{Arnett1982} presented a semianalytic model wherein the radioactive decay of $^{56}$Ni $\rightarrow ^{56}$Co $\rightarrow ^{56}$Fe emits gamma photons that diffuse through an expanding spherically symmetric and homologous SN ejecta and escape near the photosphere. The output luminosity is given as \citep[][Equation 5]{NichollEtal2017}: 

\begin{equation}\label{eq:LumOutput}
    L(t) = e^{-(t/\tau_{diff})^{-2}} \big( 1-e^{-(t/t_{leak})^{2}} \big) \int_{0}^{t} 2\text{P}(t')\dfrac{t'}{\tau_{diff}}e^{(t'/\tau_{diff})^{2}} \,\dfrac{dt'}{\tau_{diff}} 
\end{equation}
where P(t') is the source input power, a function of the SN's $^{56}$Ni mass, as shown in \citet[App. A]{CanoEtal2017}; t$_{leak}$ is a characteristic time parameter for the eventual leakage of gamma photons \citep{SollermanEtal1998}; and $\tau_{diff}$ is the effective diffusion time of the SN,

\begin{equation}\label{eq:TauDiff}
\tau_{diff} \equiv \bigg( \dfrac{2\kappa}{\beta c} \dfrac{\text{M}_{ej}}{\text{v}_{ej}}\bigg) ^{\frac{1}{2}} ,
\end{equation}
where $\kappa$ is the opacity (e.g., 0.2 cm$^{2}$g$^{-1}$ assuming only electron scattering in a hydrogen free environment) and $\beta$ is an integration constant of the source density profile ($\approx13.8$).

We use the \texttt{lmfit.minimize}\footnote{\url{https://lmfit.github.io/lmfit-py/fitting.html}} algorithm's least square method to fit critical parameters (M$_{ej}$, M$_{^{56}Ni}$ and t$_{leak}$). The best fit Arnett model is shown in Fig.~\ref{ModelFits}. The resulting parameter values for the Arnett radioactive source model are: M$_{ej}=11\pm{1}$~M$_{\odot}$, M$_{^{56}Ni}=34\pm{1}$~M$_{\odot}$, v$_{ej}=1\times10^{4}$ km~s$^{-1}$ (see Section~\ref{PCygniVelocities}), E$_{k}=6\pm{1}\times10^{51}$ erg, t$_{leak}=160\pm{4}$ days and the diffusion timescale (see Equation~\ref{eq:TauDiff}) is 54 days.  Since the total ejecta mass includes the mass of the $^{56}$Ni, the parameter results of M$_{ej}=11$~M$_{\odot}$ and M$_{^{56}Ni}=34$~M$_{\odot}$ are therefore unphysical. We note that reparametrizing the $^{56}$Ni mass as a fraction of the ejecta mass, so that the result is forced to be physical, converges to a poor fit (see Fig.~\ref{ModelFits}).

\subsection{Magnetar source model} \label{Magnetar}
In this subsection we use a maximum likelihood method, using the least squares technique, to fit the bolometric light curve. We then compare the results to a fit of the observed multi-band light curves using a Markov Chain Monte Carlo (MCMC) technique to analyze how well a magnetar can power SN\,2020qlb. 

\subsubsection{Fit to bolometric light curve} \label{Magnetar_LeastSquares}
Assuming that the central power source of the SN is from the dipole spindown energy deposition of a magnetar, \citet{InserraEtal2013} proposed that the power function P(t') to be used in Equation~\ref{eq:LumOutput} should be as follows:
\begin{equation}\label{eq:MagnetarPower}
\text{P(t')}=4.9\times 10^{46}\text{B}_{14}^{2}\text{P}_{\rm{ms}}^{-4}\dfrac{1}{(1+t'/\tau_{\rm{p}})^{2}} \quad \text{erg \ s}^{-1},
\end{equation}
where B$_{14}$ is the dipolar magnetic field strength in $10^{14}$ G, P$_{ms}$ is the initial spin period in milliseconds, and the magnetar spin-down timescale $\tau_{\rm{p}}$ is given as follows: 

\begin{equation}\label{eq:tauSpinTimeMag}
\tau_{\rm{p}}=4.7 \times \text{B}_{14}^{-2} \text{P}_{ms}^2  \quad \text{days}.
\end{equation}
The diffusion timescale $\tau_{diff}$ can be rewritten as
\begin{equation}\label{eq:diffusionTimeMag}
\tau_{diff}=1.05 \times \bigg(\dfrac{\kappa}{\beta c}\bigg)^{1/2}\text{M}_{ej}^{3/4}\text{E}_{k}^{-1/4}   \quad \text{s},
\end{equation}
where the kinetic energy of the SN ejecta E$_{k}$ is estimated assuming a homologous and spherically symmetric ejecta, with constant density, as 
\begin{equation}\label{eq:KEejecta}
\text{E}_{k}=\frac{3}{10}\text{M}_{ej}\text{v}_{ej}^{2},
\end{equation}
where v$_{ej}$ is the maximum ejecta velocity.

\begin{figure}
        \centering
        \includegraphics[width=\columnwidth]{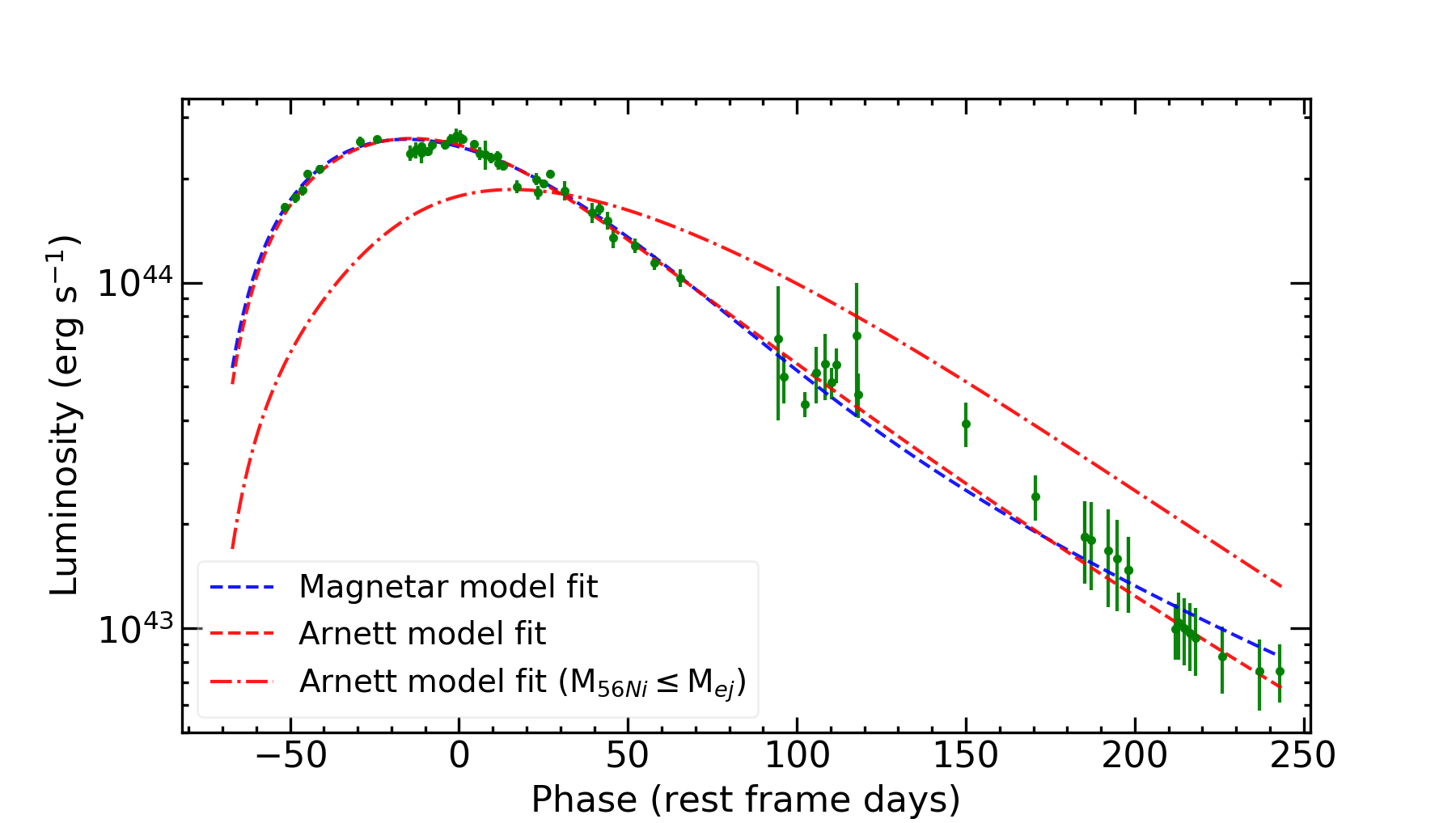}
        \caption{\label{ModelFits}Bolometric light curve plotted together with the best-fit magnetar and Arnett radioactive source models. Forcing the $^{56}$Ni mass to be less than or equal to the ejecta mass results in the red dashdot line.}
\end{figure}

Given a set of reasonable parameter (uniform) priors we fit the magnetar model using the \texttt{lmfit.minimize} least squares method to determine the best fit parameter values to be M$_{ej}=30\pm2$ M$_{\odot}$, B$_{14}=0.88\pm0.03$, P$_{\rm{ms}}=1.4\pm0.1$, v$_{ej}=1\times10^{4}$ km~s$^{-1}$ (see Section~\ref{PCygniVelocities}), t$_{leak}=309$ days, and the diffusion timescale (see Equation~\ref{eq:diffusionTimeMag}) is 86 days. The resulting best fit parameter model is plotted together with the bolometric light curve in Fig.~\ref{ModelFits}.

\subsubsection{Fit to multiband data}\label{Magnetar_MCMC}

\begin{figure*}
\centering
\includegraphics[width=0.9\textwidth,angle=0]{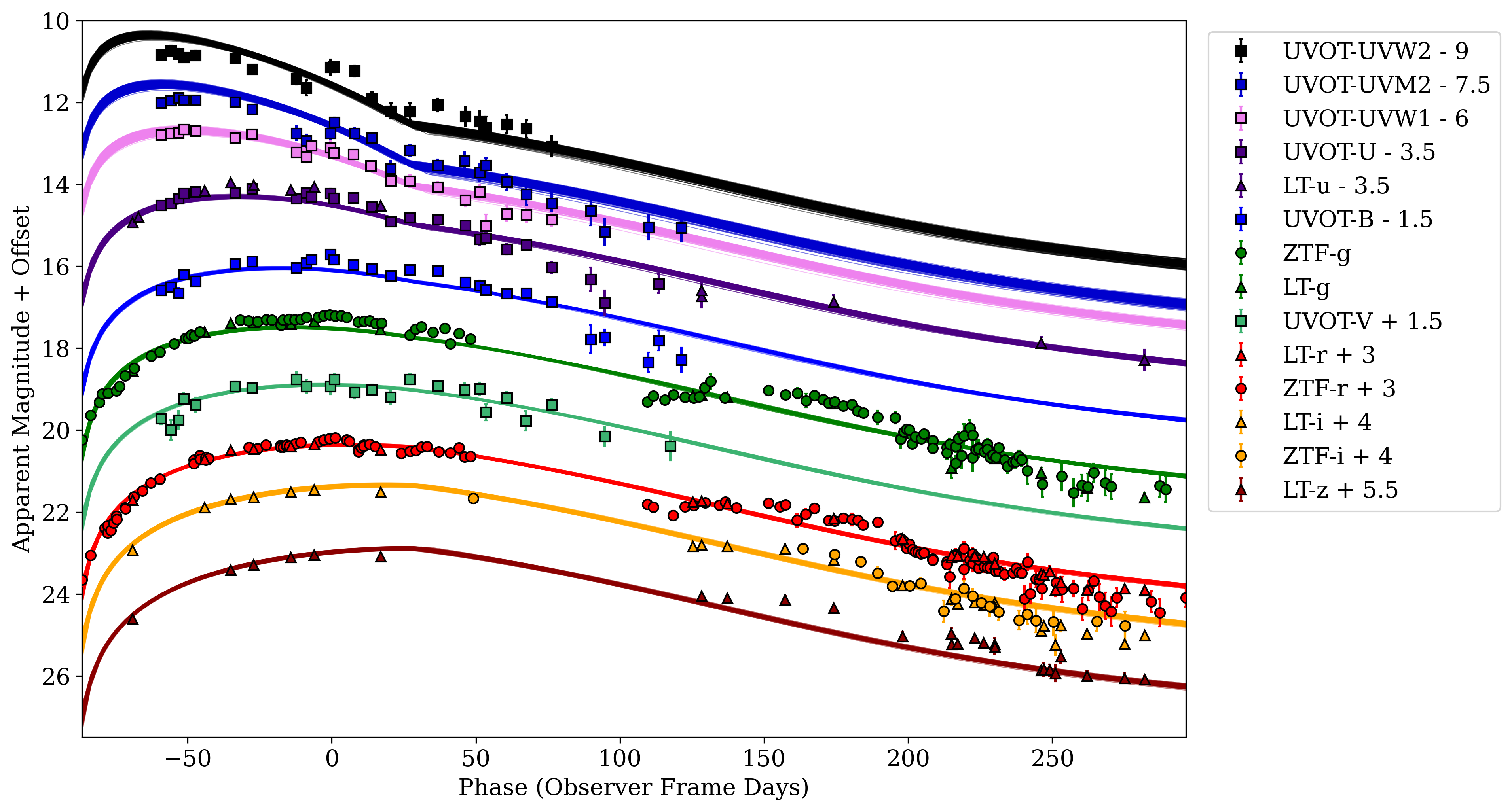}
\caption{Multi-band light curve of SN\,2020qlb inferred from the magnetar-model, with bands offset for clarity.  The colored lines show} the range of most likely models generated by \texttt{MOSFiT}. Phase $=0$ is where the ZTF $g$-band is maximized in the observer frame (MJD $59140.0$).
\label{Mosfitfit}
\end{figure*}

\begin{figure*}
\centering
\includegraphics[width=1.0\textwidth,angle=0]{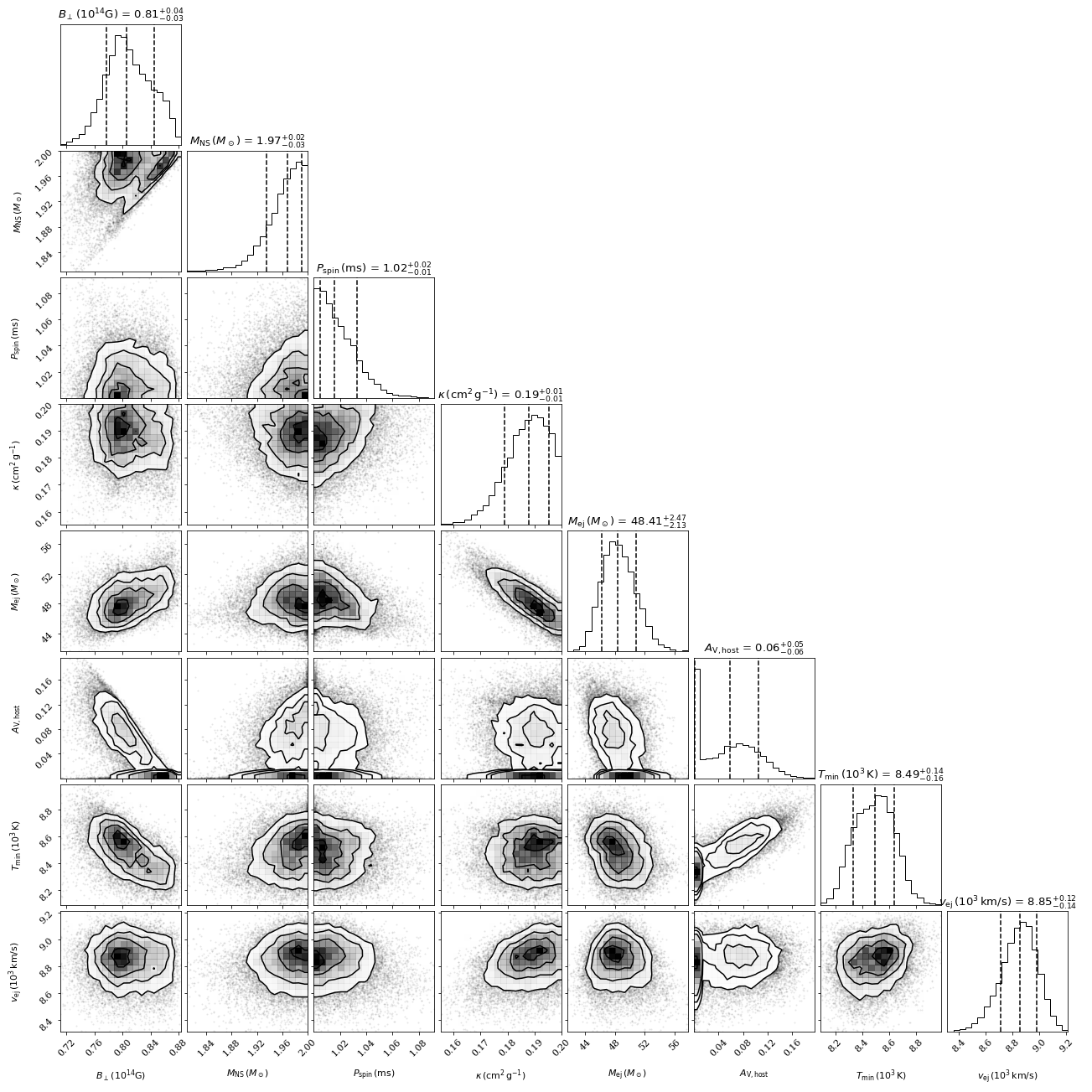}
\caption{1D and 2D posterior distributions of the
magnetar model parameters from the MOSFiT model. Median and 1$\sigma$ values are marked and labeled. These are used as the best fit values.} 
\label{Mosfitpost}
\end{figure*}

We also employed an alternative method to estimate magnetar model parameters called the Modular Open Source Fitter for Transients code (MOSFiT) \citep{GuillochonEtal2018,NichollEtal2017}. After inputting the SN redshift, all observer frame light curve data, filter information as well as reasonable priors, MOSFiT runs a MCMC process to determine the posterior distributions for 12 different parameters pertinent to a magnetar power source for the SLSN. We present the relevant results of MOSFiT for SN\,2020qlb in Table~\ref{tab:MagnetarParameters} together with the results from Sect.~\ref{Magnetar_LeastSquares}.  The MOSFiT light curve fit is shown is Fig.~\ref{Mosfitfit} and the posteriors are shown in Fig.~\ref{Mosfitpost}.

\begin{table}
\caption{\label{tab:MagnetarParameters} Comparison of two methods for magnetar modeling and their best fit (median and 1$\sigma$) parameters. MOSFiT priors are also shown where U is for uniform and G is for Gaussian.  $^{*}$ indicates that the value was fixed in the model run; and NA stands for not applicable.}
\centering
\begin{tabular}{lllc}
\hline
\hline
 Parameter & MOSFiT priors & MOSFiT  & Least squares  \\ \hline
 M$_{ej}$ [M$_{\odot}$] & [0.1, 100, log-U] & $48.4_{-2.1}^{+2.5}$  & 30$\pm2$ \\
 M$_{NS}$ [M$_{\odot}$] & [1, 2, U] & $1.97_{-0.03}^{+0.02}$  &  NA\\
 P$_{ms}$ & [1, 10, U] & $1.02_{-0.01}^{+0.02}$ & 1.4$\pm0.1$ \\
 B$_{14}$ [$10^{14}$G] & [0.1, 10, U] & $0.81_{-0.03}^{+0.04}$ & 0.88$\pm0.03$ \\
 $\kappa$ [cm$^{2}$g$^{-1}$]  & [0.05, 0.2, U] & $0.19_{-0.01}^{+0.01}$ & $0.2^{*}$ \\
 v$_{ej}$ [km~s$^{-1}$] & [10$^3$, 10$^5$, G] & $8854_{-144}^{+124}$ & $10000^{*}$ \\
 T$_{min}$ [K] & [$3\times10^{3}$, 10$^4$, U]  & $8493_{-158}^{+144}$ & NA \\
 A$_{V,host}$ & [10$^{-5}$, 10$^2$, log-U] & $0.06_{-0.06}^{+0.05}$ & $0^{*}$ \\
\end{tabular}
\end{table}

We find good agreement between the values of A$_{V,host}$, $\kappa$, and v$_{ej}$ assumed for our least squares fit and those found by MOSFiT, as well as between T$_{min}$ found by MOSFiT and the late-time photospheric temperature calculated in Sect.~\ref{Blackbodyfits} (see Fig.~\ref{TempEv}).  We also find the posteriors of both M$_{NS}$ and P$_{\rm spin}$ peak close to the edges of their respective priors, which are informed by conservative estimates of the Tolman-Oppenheimer-Volkoff (TOV) limit and mass-shedding limit respectively. This means that we herein find this system to contain a magnetar close to its maximum mass spinning at close to its breakup velocity.

Both the MOSFiT and least squares methods estimate the kinetic energy using Equation~\ref{eq:KEejecta} to be approximately $2\times 10^{52}$ erg.  

The total amount of rotational energy stored in a magnetar is estimated by \citet{Kasen2017} as
\begin{equation}\label{eq:RotationalEnergy}
    \text{E}_{\rm{rot}} \approx 2.5\times10^{52}\text{P}_{\rm{ms}}^{-2}\Big(\frac{\text{M}_{NS}}{1.4\text{M}_{\odot}}\Big)^{3/2} \
    \quad \text{erg.}
\end{equation}
The MOSFiT method thereby estimates the available rotational energy to power the SN light curve to be $4.0\times 10^{52}$ erg while the least squares method (with M$_{NS}=1.97$ M$_{\odot}$) only predicts it to be $2.1\times 10^{52}$ erg. However, both of these values are well above the total integrated light curve radiated energy of $2.8\pm{0.3}\times 10^{51}$ erg (see Sect.~\ref{BoloInterpol}). 

The magnetar spin-down timescales calculated via the Least-Squares and MOSFiT methods are 3.8 and 4.0 days respectively, while the diffusion timescales are 86 and 82 days respectively. \cite{SuzukiMaeda21} show that if the spin-down and diffusion timescales are comparable, then a large fraction of the rotational energy is expected to be contributed to the SN luminosity, which is typical for SLSNe. If the spin-down timescale is much shorter than the diffusion timescale then most of the energy is expected to be contributed to the kinetic energy, which is more typical for SN Ic-BL. From supernova surveys and modeling efforts, the average SLSN-I rise time (which is roughly the diffusion time) is $\sim$ 40 days \citep{ChenEtal2022a} and spin-down timescale is $\sim$ 15 days \citep{NichollEtal2017, ChenEtal2022b}, a factor of $\sim$ 2-3; while the average rise time of a SN Ic-BL is $\sim$ 15 days \citep{Taddia2019}, and they are usually modeled with spin-down timescales of $\sim$ an hour \citep{SuzukiMaeda21}, a factor of $>$ 100.  Both fitting methods give $\tau_{\rm diff}/\tau_{\rm p} \sim$ 20 for SN\,2020qlb, which is intermediate between typical SLSNe-I and SNe Ic-BL, and is consistent with a large fraction of the rotational energy being converted to kinetic energy in this SN.

\subsubsection{Predicted Radio and Soft X-ray Counterparts} \label{radxray}

Radio observations can provide an interesting clue as to the nature of the supernova power source, as both a magnetar engine \citep{2016MNRAS.461.1498M, 2018MNRAS.474..573O} and CSM interaction \citep{2018SSRv..214...27C} can produce radio emission, but on different timescales and with different spectra.  Three SLSNe-I have already been detected in radio: PTF10hgi \citep{2019ApJ...876L..10E, 2019ApJ...886...24L, 2020MNRAS.498.3863M, 2021ApJ...911L...1H}, which is consistent with the magnetar model; SN\,2017ens \citep{2021ATel14393....1C}, which transitioned from BL-Ic to SLSN and is likely powered by a combination of a magnetar and CSM interaction \citep{Chen2018}; and SN\,2020tcw \citep{2021ATel14418....1C}, which was detected only a few months after explosion and is likely due to a CSM interaction.

\begin{figure}
        \centering
        \includegraphics[width=\columnwidth]{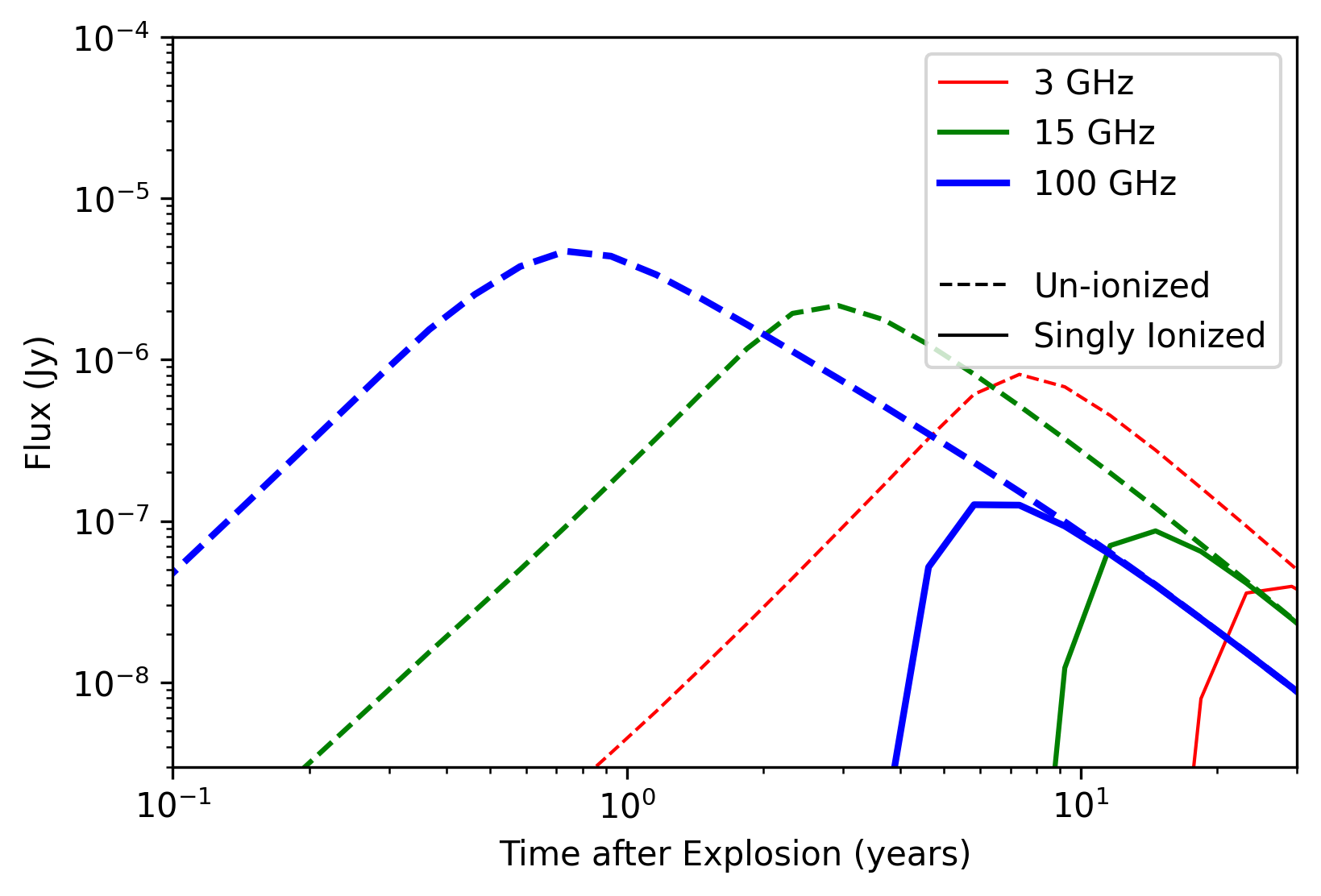}
        \caption{Predicted radio emission for SN\,2020qlb at 3 (red), 15 (green), and 100 (blue) GHz.  The dashed and solid lines represent estimates of free-free absorption using an un-ionized and singly ionized oxygen ejecta, respectively.}
        \label{radio}
\end{figure}

We use the magnetar parameters found from the MOSFiT models in Sect.~\ref{Magnetar_MCMC} to calculate the expected radio emission for SN\,2020qlb.  We use the model previously presented in \citet{2018MNRAS.474..573O, 2019ApJ...886...24L, 2021ApJ...912...21E}, which assumes pulsar wind nebula (PWN) microphysics calibrated to the Crab Nebula \citep{2010ApJ...715.1248T, 2013MNRAS.429.2945T}, which has a broken power-law electron injection spectrum with an injection Lorentz factor $\gamma_b = 6 \times 10^5$ and spectral indices $q_1 = 1.5$ and $q_2 = 2.5$.  In Fig.~\ref{radio} we show the predicted light curves at 3, 15, and 100 GHz, which correspond to VLA bands S and Ku and ALMA band 3, respectively. Since the ionization state of the ejecta is not fully understood, we use free-free absorption estimates based on an un-ionized (dashed lines) and fully singly-ionized (solid lines) ejecta as two extreme cases.

Previous observations with VLA and ALMA \citep[e.g.,][]{2019ApJ...886...24L, 2021ApJ...912...21E, 2021MNRAS.508...44M} have usually had noise levels of about 10 $\mu$Jy, while SN\,2020qlb is only expected to reach $\sim$ 1 $\mu$Jy for an un-ionized ejecta and 0.1 $\mu$Jy for a singly ionized ejecta, so it is likely not a good candidate for radio follow-up.  Additionally, changing the microphysics of the PWN could result in either a low-magnetization Compton-dominated nebula or a high-magnetization synchrotron-dominated nebula \citep{VurmMetzger2021, 2021MNRAS.508...44M}, which both result in a decrease in radio emission and an increase in gamma-ray emission, making the object even harder to detect.

For soft X-rays, the main absorption process at 0.3-10 keV is photoelectric absorption which has an optical depth that can be expressed as $\tau = K\rho R$. $\rho$ is the ejecta density, $R$ is the ejecta radius, and $K$ is a mass attenuation coefficient that can be estimated as 2.5 cm$^2$ g$^{-1}$ $(Z/6)^3 (E_\gamma / $10 keV$)^{-3}$, where $Z$ is the average atomic number and $E_\gamma$ is the photon energy \citep{Muraseetal, Kashiyamaetal}.  For SN\,2020qlb, we find $\tau \approx 5 \times 10^{6} (t/$day$)^{-2} (E_\gamma / $10 keV$)^{-3}$, so soft X-rays are not expected to be able to escape for $\gtrsim$ 5 years.  The X-rays can also escape if the ejecta is completely ionized \citep{MetzgerEtal2014}, but that is unlikely to happen for SLSNe with more massive ejecta, such as SN\,2020qlb.

\section{Light curve undulations} \label{Undulations}
One striking feature of SN\,2020qlb is its light curve undulations. Despite similar findings in other SLSNe \citep[e.g.,][]{InserraEtal2017,NichollEtal2016}, the physical mechanisms behind the so-called bumps are not well understood. In this section, we analyze the light curve undulations present in SN\,2020qlb (amplitudes, timescales); their interpretation is discussed in Section~\ref{UndulationSource}.

\begin{figure*}
        \centering
        \includegraphics[width=\textwidth]{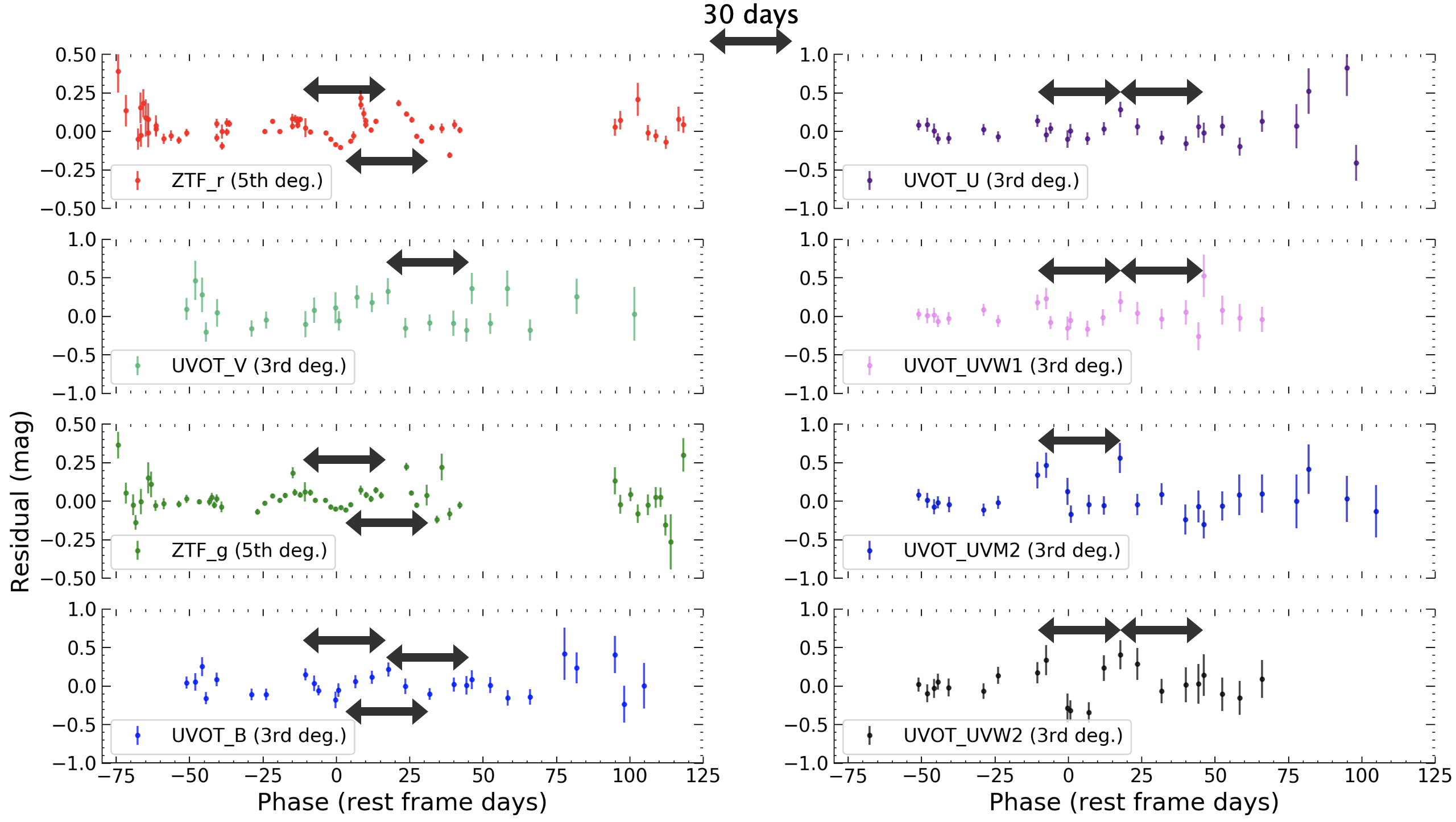}
        \caption{\label{UndulationsAllFilters} Light curve residuals after subtracting a polynomial (order indicated in each panel) to the first 200 days of data. Visual inspection indicates fluctuations with an approximate timescale of $\sim$30 days in all filters; this is indicated in each subplot with an arrow showing time between peaks/troughs. }
\end{figure*}

To characterize the undulations, we first subtracted from each filter a polynomial fit capturing the large-scale shape of the light curves. We chose the lowest-order polynomial that still fits the overall shape; for most filters a third-degree polynomial was sufficient, while the two ZTF filters required a fifth-order one. Fig.~\ref{UndulationsAllFilters} shows the resulting residual light curves.

Visual inspection of each subplot in Fig.~\ref{UndulationsAllFilters} shows residual fluctuations in all filters. The typical timescale (peak to peak or trough to trough) appears to be about 30 days; these are highlighted with arrows on the figure. The fact that the arrows generally match across filters suggests that the undulations are approximately monochromatic.

\begin{figure}
        \centering
        \includegraphics[width=\columnwidth]{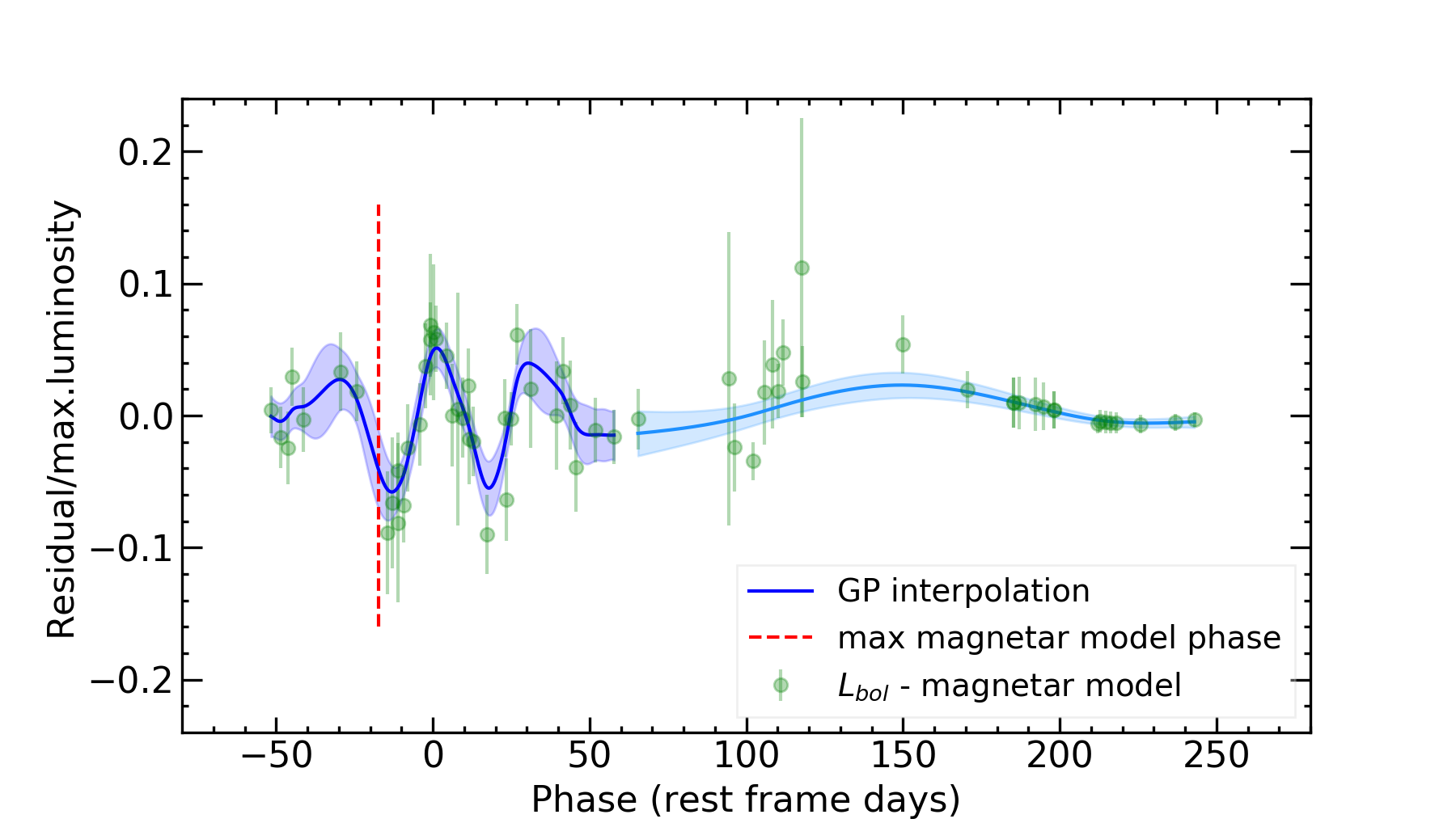}
        \caption{\label{GPmagnetarresidual}Bolometric light curve less magnetar residual and the GP interpolation of the residual, both normalized by the maximum magnetar model luminosity. The phase of the magnetar model maximum (-19.4 days) is marked with a vertical line for reference.}
\end{figure}

To better quantify the timescales of the undulations, we analyzed the bolometric light curve. At each of the phases we calculated the residual between the bolometric light curve and the best fit magnetar model. The least squares method method was selected since it utilized the constructed bolometric light curve to identify the best magnetar model fit. The resulting residual is shown in Fig.~\ref{GPmagnetarresidual}; visual inspection suggests an oscillatory appearance for the first 150 days. A GP interpolation of the residual is also shown in dark blue. The center maximum, of the three GP interpolation maxima, matches the peak M$_{g}$ phase indicating that the undulation is included in the maximum brightness determination. We also note that the peak of the magnetar model (marked in orange) is at a phase of -19.4 days.

\begin{figure}
        \centering
        \includegraphics[width=\columnwidth]{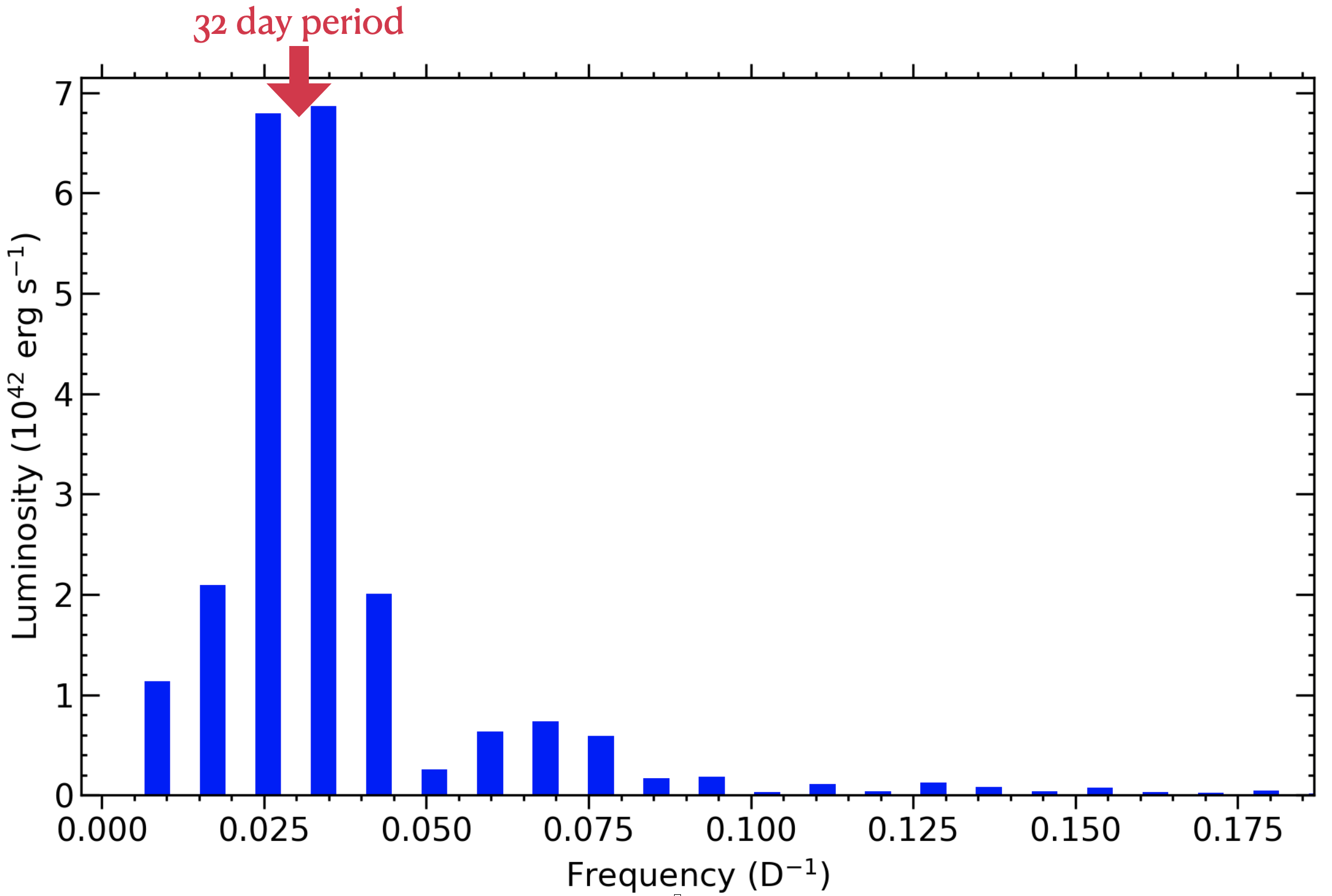}
        \caption{\label{FFT_GP_magnetarresidual}Fast Fourier transform of the GP interpolation of the bolometric LC less magnetar model residual for phases up to +60 days} as shown in darker blue in Fig.~\ref{GPmagnetarresidual}.
\end{figure}

In order to identify the timescale of the undulations in the GP interpolated residual we performed a discrete fast Fourier transform (FFT) using the $\texttt{scipy.fftpack.fft}$\footnote{\url{https://docs.scipy.org/doc/scipy/reference/generated/scipy.fftpack.fft.html}} algorithm to create a periodogram of the underlying frequencies in the residual. Fig.~\ref{FFT_GP_magnetarresidual} shows that a timescale of approximately $32\pm6$ days is present in the residual, which matches well what is seen by eye in the individual filter residuals in Fig.~\ref{UndulationsAllFilters}.

The energy scale of the undulations is estimated by the residual's maximum amplitude of approximately $1.7\times10^{43}$ erg/s. This is about 6$\%$ of the peak bolometric luminosity of $2.6\times10^{44}$ erg/s.

We also observe an additional bump in the residual between phases of +100 and +180 days in Fig.~\ref{GPmagnetarresidual}. A GP interpolation for this time period is shown in light blue as well. This late-time bump appears to have a lower amplitude and a longer timescale than the earlier ones.

Light curve bumps have been discovered in many other SLSNe-I. \citet{HosseinzadehEtal2021} found bumps or undulations in 44-76\% of the post peak light curves of 34 SLSNe. Similarly, \citet{ChenEtal2022a} observed that 39\% - 66\% of 77 SLSNe-I from the ZTF-I operation had undulations with an average duration of 21 days and an average amplitude of 4\% of maximum brightness.

\section{Host Galaxy} \label{sec:host}

As seen in Fig.~\ref{HostGalaxy_Image}, the host galaxy of SN\,2020qlb appears to be a faint, blue dwarf galaxy.
In this section, we discuss the properties of the galaxy in more detail, and put it in context of the population of SLSN-I host galaxies.

\subsection{HST image and morphology}
\begin{figure}
    \centering
    \includegraphics{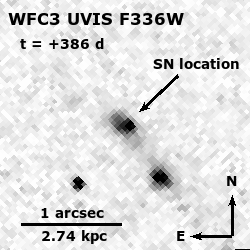}
    \caption{\textit{HST} image of the host galaxy of SN\,2020qlb, corresponding to a rest-frame wavelength of 2900~\AA, and taken at a phase of +386 days past the g-band peak. The host galaxy appears to have two knots of emission, with the supernova located at the northern knot. The apparent point source could include supernova emission, or be a star-forming region in the host galaxy; a template image would be needed to tell the two apart. The source visible in the bottom left of the image is caused by a cosmic ray.}
    \label{fig:hst_image}
\end{figure}

SN\,2020qlb was observed by the \textit{Hubble Space Telescope} (\textit{HST}) WFC3/UVIS on January 7, 2022, as part of Snapshot program 16657 (PI Fremling), corresponding to a phase of +367 rest-frame days past g-band maximum. The image in Fig.~\ref{fig:hst_image} is taken in F336W, with a corresponding rest-frame effective wavelength of 2900~\AA. We see two bright knots of emission, with the supernova location corresponding to the northern knot as indicated by the arrow. The total systematic astrometric uncertainty, dominated by the ZTF, is about 0.1 arcsec. With more stars in the UVIS image we could have made a more precise astronometric matching to better determine the exact supernova location within the galaxy. We note that this kind of morphology, which could be either an interacting system or a dwarf galaxy with multiple regions of strong star formation, is not unusual among SLSN-I host galaxies \citep{LunnanEtal2015,OrumEtal2020}.

We also note that the northern knot appears, by eye, to consist of a point source on top of more extended emission. This point source could be UV light from the supernova still visible at +367 days; PSF photometry yields an apparent magnitude $m_{2900\AA} = 25.06 \pm 0.05~{\rm mag}$, which corresponds to an absolute magnitude of $M_{2900\AA} = -14.47 \pm 0.05~{\rm mag}$. However, to truly ascertain whether this is supernova light or simply a brighter region within the host galaxy would require a second epoch of \textit{HST} imaging in order to do proper host subtraction.

\subsection{Emission line diagnostics} \label{sec:EmissionLineDiag}
The Keck spectrum taken at a phase $+461$~days past maximum (see Fig.~\ref{SpectralEv}) is dominated by host galaxy light, and contains a wealth of emission lines that can be used to analyze the properties of the underlying H~II region. We measure emission line fluxes by fitting Gaussian profiles to the (generally unresolved) host galaxy lines; the results are listed in Table~\ref{tab:line_fluxes}.

\begin{table}
\centering
\small
\caption{\label{tab:line_fluxes} Observed host galaxy emission line fluxes, uncorrected for MW dust extinction.}
\begin{tabular}{cc}
\hline
\hline
Line         & Flux       \\
           &  $10^{-16}~{\rm erg~s}^{-1}{\rm cm}^{-2}$  \\ \hline
$[$S II$]\lambda6731$ &  $0.91 \pm 0.08$ \\
$[$S II$]\lambda6717$ &  $1.27 \pm 0.10$ \\
$[$N II$]\lambda6584$ &   $1.16 \pm 0.23$ \\ 
H$\alpha$ & $19.28 \pm 0.33$ \\
$[$O III$]\lambda5007$ &  $28.04 \pm 0.25$ \\
$[$O III$]\lambda4959$ &  $9.21 \pm 0.19$ \\
H$\beta$ & $5.66 \pm 0.28$ \\
$[$O III$]\lambda4363$ &  $0.41 \pm 0.15$ \\
H$\gamma$ &  $2.33 \pm 0.15$ \\
H$\delta$ & $0.98 \pm 0.15$ \\
H$\epsilon$ + [Ne III]$\lambda3968$ & $1.22 \pm 0.15$ \\
H$\zeta$ & $0.67 \pm 0.13$ \\
$[$Ne III$]\lambda3869$ &  $2.26 \pm 0.16$ \\
$[$O II$]\lambda3727$ &  $9.88 \pm 0.22$ \\ \hline
\end{tabular}
\end{table}

After correcting for Milky Way extinction, we measure a Balmer decrement ${\rm H}\alpha / {\rm H}\beta = 3.20 \pm 0.16$, indicating moderate host galaxy extinction. Assuming intrinsic ratios corresponding to Case B recombination \citep{Osterbrock1989}, we derive a contribution $E(B-V)_{\rm host} = 0.10 \pm 0.05~{\rm mag}$.

We marginally detect the auroral [O\,III]$\lambda$4363 line, allowing for the electron temperature to be calculated and the oxygen abundance to be measured directly. We use the Python package {\tt PyNeb} \citep{PyNeb} to iteratively calculate the O$^{++}$ electron temperature and the electron density $n_e$ from the ratios of [O\,III]$\lambda$4363\,/\,[O\,III]$\lambda$5007 and [S II]$\lambda$6731\,/\,[S II]$\lambda$6717, respectively. The O$^{+}$ electron temperature is then obtained assuming the relation 
\begin{equation}
    T_e({\rm O}^{+}) = 0.7 \times T_e({\rm O}^{++}) + 0.3,
\end{equation} 
where $T_e({\rm O}^{+})$ and $T_e({\rm O}^{++})$ are in units of 10000~K \citep{CampbellEtal1986}. We then use the ratios of [O III]$\lambda$5007\,/\,[O\,III]$\lambda$4959\,and [O II]$\lambda$3727\,/\,H$\beta$ to calculate the O$^{++}$/H and O$^{+}$/H abundances, respectively, again using {\tt PyNeb}. The final oxygen abundance is obtained from summing these two contributions. Using a Monte Carlo approach to resample the fluxes within their errors to calculate the uncertainty, we obtain a final metallicity of $12 + \log({\rm O/H}) = 8.0 \pm 0.2 ~{\rm dex}$. Taking the solar oxygen abundance to be $12 + \log({\rm O/H}) = 8.69 \pm 0.05 ~{\rm dex}$ \citep{AsplundEtal2009}, this corresponds to a metallicity of $Z \simeq 0.2 Z_{\odot}$.

Using the extinction-corrected H$\alpha$ flux, we can also calculate a star formation rate using the relation ${\rm SFR} (M_{\odot}{\rm yr}^{-1})= 7.9 \times 10^{-42} \times L\left(\rm H\alpha\right) {\rm erg~s}^{-1}$ \citep{Kennicutt1998}. This yields a star formation rate of $1.4~M_{\odot}{\rm yr}^{-1}$ for the host galaxy of SN\,2020qlb. We note that this estimate is based on a spectrum ($+461$~days) that did not have contemporaneous calibration photometry but was calibrated using host galaxy photometry.

\subsection{Host SED modeling}
\begin{figure}
    \centering
    \includegraphics[width=0.48\textwidth]{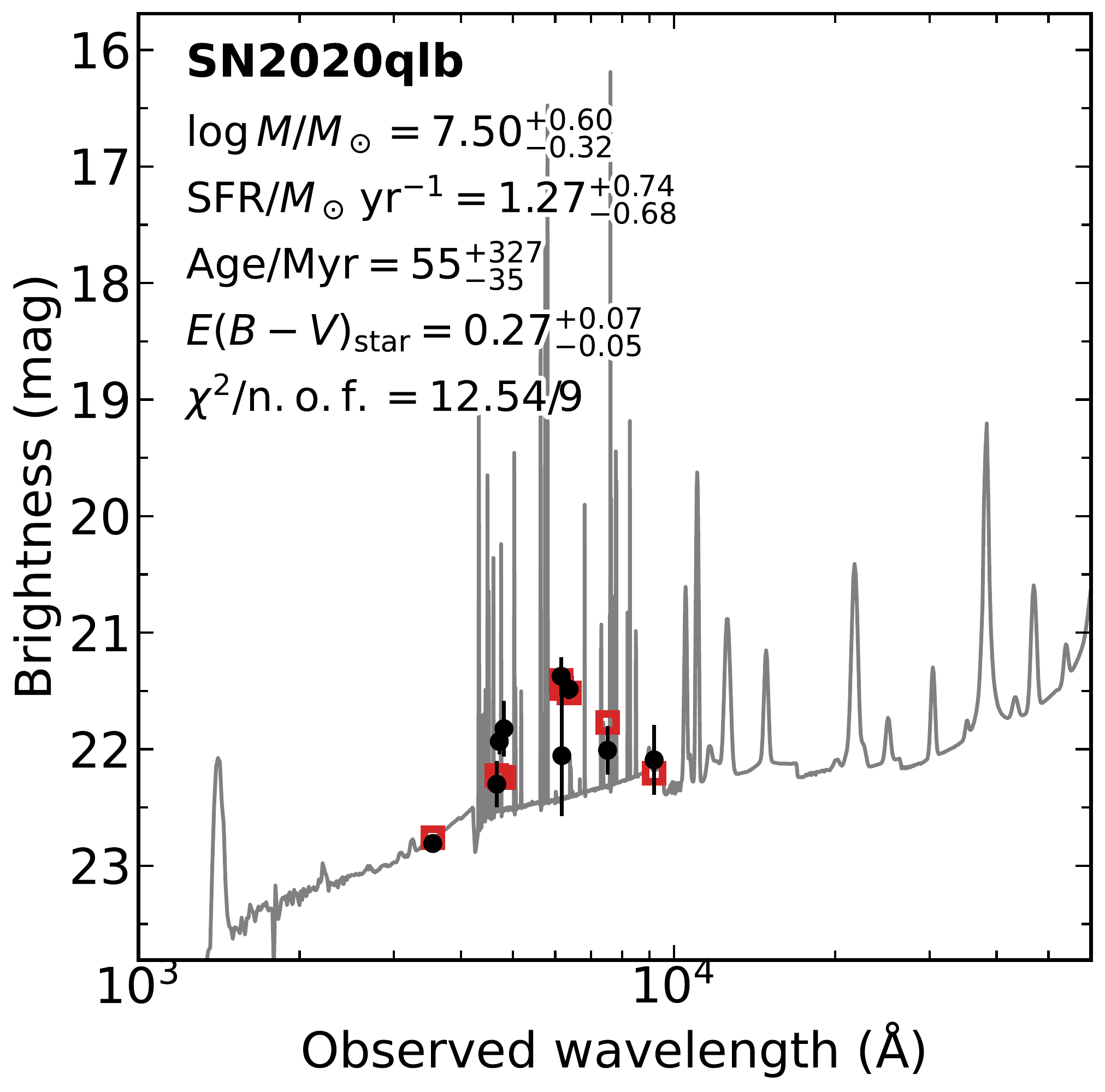}
    \caption{Spectral energy distribution (SED) of the SN\,2020qlb host galaxy from 1000 to 60000~\AA\ (black data points, Table~\ref{tab:hostphot}). The solid line displays the best-fitting model of the SED. The red squares represent the model-predicted magnitudes. The fitting parameters are shown in the upper-left corner. The abbreviation ``n.o.f.'' stands for numbers of filters.}
    \label{fig:gal_sed}
\end{figure}

Figure \ref{fig:gal_sed} shows the observed host galaxy SED from 3000 to 10000~\AA. We modeled the SED with the software package {\tt Prospector} version 1.1 \citep{Leja2017a} which uses the Flexible Stellar Population Synthesis (FSPS) code \citep{Conroy2009a} to generate the underlying physical model and python-fsps \citep{ForemanMackey2014a} to interface with FSPS in Python. The FSPS code also accounts for the contribution from the diffuse gas (e.g., \ion{H}{ii} regions) based on the Cloudy models from \citet{Byler2017a}. Furthermore, we assumed a Chabrier initial mass function \citep{Chabrier2003a} and approximated the star formation history (SFH) by a linearly increasing SFH at early times followed by an exponential decline at late times (functional form $t \times \exp\left(-t/\tau\right)$). The model was attenuated with the \citet{Calzetti2000a} model.

The best fit, shown in gray in Fig.~\ref{fig:gal_sed}, suggests a low-mass star-forming galaxy with a mass of $7.50^{+0.60}_{-0.32}~M_\odot$ and a star formation rate of $1.27^{+0.74}_{-0.68}~M_\odot\,{\rm yr}^{-1}$. The mass and the star formation rate are in the expected parameter space of host galaxies of SLSNe-I at similar redshifts \citep{PerleyEtal2016, SchulzeEtal2018} albeit in the lower half. The attenuation inferred from the SED modeling is broadly consistent with what is obtained from the Balmer decrement (Sect. \ref{sec:EmissionLineDiag}).

\section{Discussion} \label{Discussion}
In this section we begin by discussing and comparing SN\,2020qlb to the unique criteria and general characteristics of SLSNe-I as described by \citet{Howell2017} and by \citet{Gal-Yam2019}. Potential light curve power sources are then discussed, followed by a review of possible undulation power sources.

\subsection{SLSN-I concordance} \label{Phases}

In this subsection we discuss the distinctive characteristics of SLSNe-I based on their four phases as presented by \citet{Gal-Yam2019}: 1. Early bump, 2. Hot photosphere, 3. Cool photosphere, 4. Nebular. We therein discuss how SN\,2020qlb compares to each typical property. 

\citet{ChenEtal2022b} found an early bump in 3/15 (6-44\% with confidence limit of 95\%) SLSNe-I from their ZTF Phase-I survey with at least four epochs of prepeak photometry. SN\,2020qlb's lack of an early light curve bump is therefore not unusual.

SLSN rise times, that is to say from explosion to the luminosity peak, typically range from $\sim$20 to $>$100 days in the rest frame of the SN \citep{Gal-Yam2019}. SLSN light curve rise times from 1/e maximum to peak are typically between $\sim$15 and $>$60 days (see Fig.~\ref{MagGvsRiseTime}). SN\,2020qlb's 77.1 day rise time from explosion to peak, although on the longer side, is therefore typical for a SLSN-I (Fig.~\ref{MagGvsRiseTime}).

Similarly, the peak luminosity of SN\,2020qlb is shown on Fig.~\ref{MagGvsRiseTime} compared to the ZTF-I sample of \citet{ChenEtal2022a}. With a peak g-band absolute magnitude of M$_{g}=-22.25\pm0.01$~mag, SN\,2020qlb is in the upper range of typical SLSNe, and well above any threshold to be considered superluminous \citep{QuimbyEtal2018,Gal-Yam2019}. 

The hot photospheric phase, which includes the peak, is characterized by a hot (blue) spectral continuum with decreasing blackbody temperatures of up to 20000~K, which is indeed what we find for SN\,2020qlb in Fig.~\ref{TempEv}. Several absorption features (O\,I, O\,II and C\,II) are detected on top of the continuum. In particular, O\,II absorption features in the blue part of the visible spectrum are unique to SLSNe-I and are found persistently prior to the peak \citep{Gal-Yam2019}. As discussed in Sect.~\ref{SpectralClassification}, SN\,2020qlb has the typical O\,II "W" feature near 4500Å in its early spectra. Expansion velocities derived from the P-Cygni line profiles of O\,I, O\,II, Fe\,II et al. during the hot photospheric phase are typically estimated to be between 10000 and 15000 km~s$^{-1}$ \citep{QuimbyEtal2018,Gal-Yam2019}. In Fig.~\ref{VelocityEv}, SN\,2020qlb has an early velocity of $\sim$10000 km~s$^{-1}$.

During the cool photospheric phase the photosphere cools and expands while the spectrum evolves to resemble typical Type Ic SN spectra. Meanwhile the unique O\,II features from the hot photospheric phase weaken as the temperature typically falls below $\sim$12000 K \citep{Gal-Yam2019}. In Sect.~\ref{SpectralClassification} we show how SN\,2020qlb's late spectra are typical for SLSNe-I.

The host galaxy of SN\,2020qlb is also quite typical of SLSN-I host galaxies, with a low mass ($\log(M/M_{\odot}) \simeq 7.5$), low metallicity ($12 + \log({\rm O/H}) \simeq 8.0$, direct method), and a high star formation rate ($1.27^{+0.74}_{-0.68}~M_\odot\,{\rm yr}^{-1}$). These are all within the typical range of SLSN-I host galaxies at this redshift, albeit at the more extreme end (e.g., \citealt{LunnanEtal2014,LeloudasEtal15,PerleyEtal2016,ChenEtal2017b,SchulzeEtal2018,SchulzeEtal2021}).

SN\,2020qlb clearly meets important criteria regarding brightness, spectral features and evolution. And since no characteristic is counter-indicative, we find that SN\,2020qlb is a typical SLSN-1.

\subsection{Light curve power source} \label{Power}

\subsubsection{Radioactive decay}

Given the extreme level of SN\,2020qlb's brightness it is possible to suspect a power source wherein a $e^{-}$/$e^{+}$ pair-production instability SN (PISN) explosion could annihilate the progenitor star completely. \citet{KasenEtal2011} show that models of stars with initial masses in the range of 140 M$_{\odot}$ to 260 M$_{\odot}$ die in thermonuclear runaway explosions resulting in the synthesis of up to 40 M$_{\odot}$ of $^{56}$Ni. The best fit radioactivity model shown in Fig.~\ref{ModelFits} estimates that $34\pm1$ M$_{\odot}$ of $^{56}$Ni is required to explain the SN\,2020qlb bolometric light curve, an amount within the PISN model prediction.

The Arnett $^{56}$Ni radioactive decay model fitting done in Sect.~\ref{RadioactiveModel} estimates M$_{ej}$, v$_{ej}$, M$_{^{56}Ni}$ and t$_{leak}$. v$_{ej}$ is normally measured from high resolution spectra (see Sect.~\ref{Blackbodyfits}). The diffusion time $\tau_{diff}$ (see Equation~\ref{eq:TauDiff}) describes the characteristic time frame for light to diffuse through the expanding ejecta and is calculated using both M$_{ej}$ and v$_{ej}$. The gamma photon leakage time t$_{leak}$ is given by \cite{ClocchiattiWheeler1997} as t$_{leak}\approx(\text{M}_{ej}/\text{E}_{51})^{1/2}(\text{M}_{ej})^{1/2}\propto\text{M}_{ej}^{1/2}/\text{v}_{ej}$. So, in essence, the Arnett radioactive decay model only has two characteristic parameters, M$_{ej}$ and M$_{^{56}Ni}$. However, for SN\,2020qlb the $^{56}$Ni mass is estimated to be significantly more than the ejecta mass (see Section~\ref{RadioactiveModel}). Since this is unphysical we discard the Arnett model describing a $^{56}$Ni radioactive source. The same argument requires us to discard the PISN radioactivity model as well.

\subsubsection{Magnetar}
The magnetar model fitting done in Sect.~\ref{Magnetar} is also able to trace the bolometric light curve. The MOSFiT MCMC method tests ranges of values for parameters that are assumed constant in the least squares fitting method. Table~\ref{tab:MagnetarParameters} indicates how the two methods compare for key parameters such as M$_{ej}$, P$_{ms}$, B$_{14}$ and v$_{ej}$. By combining M$_{ej}$ and v$_{ej}$ into one parameter E$_{KE}$, the two methods achieve better agreement. So, in essence, the magnetar model has three characteristic parameters which are capable of reproducing SN\,2020qlb's light curve, that is P$_{ms}$, B$_{14}$, and E$_{KE}$. \citet{NichollEtal2017} present MOSFiT results for 38 SLSNe-I wherein median value ranges for P$_{ms}=2.4^{+1.6}_{-1.2}$ and B$_{14}=0.8^{+1.1}_{0.6}$ as well as the total range for E$_{KE}$= 0.55 to 25.06 $\times10^{51}$ [erg] all span the results of SN\,2020qlb in Table~\ref{tab:MagnetarParameters}. The soft X-ray nondetections (see Sect.~\ref{XrayDetections} and Sect.~\ref{radxray}) are also consistent with the magnetar model. We note that radio observations (see Sect.~\ref{radxray}) could be used as a potential test of this scenario, but the predicted fluxes are too low for current radio interferometers.

The model for the smooth spin-down of a magnetar can only impart its rotational energy into the bolometric light curve in a smooth way. It therefore can only trace the general shape of the light curve and not the undulations as discussed in Sect.~\ref{Undulations}. Eventual magnetar-powered undulations not captured by the model are discussed in Sect.~\ref{CentralSource}.

Given that each of the parameter estimates are within physically possible ranges (see Sect.~\ref{Magnetar_MCMC}) the magnetar model is retained.

\subsubsection{CSM interaction}\label{CSMinteraction}

An additional potential external power source of SN light curves is the collisional interaction of the SN ejecta with circumstellar material (CSM). Strong shocks can convert the ejecta's kinetic energy into radiation energy. The CSM can potentially result from stellar winds, binary mergers or interaction, or stellar eruptions \citep[see][]{Smith2014}.

\citet{Gal-Yam2019} indicates that most CSM interacting SNe have strong and narrow emission lines as encountered in the spectra of H-rich Type IIn (n refers to "narrow"), He-rich Type Ibn, Ia-CSM and some Ic SNe. However, a lack of narrow lines does not always imply a noninteracting SN \citep[see e.g., Type II-L SN\,1979C and SLSN\,II SN\,2008es][]{FranssonEtal1984,BhirombhakdiEtal2019}. There are a few examples of CSM interaction in SLSNe-I noted by \citet{YanEtal2017}. However, as noted in Sect.~\ref{SpectralProperties}, we find no such signature spectral features for SN\,2020qlb as shown in Fig.~\ref{SpectralEv}.

\citet{Gal-Yam2019} writes that there are no existing published models employing CSM interaction power that can fit SLSN-I spectra. However, there have been efforts to fit SLSN light curves with CSM interaction models. Hybrid, for example CSM plus radioactive decay, models such as the semianalytic model \texttt{MINIM} \citep{ChatzopoulosEtal2013} employ both SN and CSM parameters in a $\chi^{2}$ minimization fit. The increased number of available parameters in CSM interaction models should enable bolometric light curves to be reproduced very well. \citet{LiuEtal2018} show that by increasing the number of CSM interactions it is possible to model even complex light curves, for example iPTF15esb $\&$ iPTF13dcc. In addition, \citet[Fig. 1]{LiuEtal2018} show how a triple ejecta CSM (in total $\sim4M_{\odot}$) interaction can fit the undulating bolometric light curve of iPTF15esb. In essence any light curve could eventually be fitted with multiple CSM interactions of different magnitudes occurring at different times.

We used MOSFiT to perform two CSM model \citep{ChatzopoulosEtal2013, VillarEtal2017, JiangEtal2020} fits using two different kinds of CSM by setting the slope of the CSM density profile to be $s = 0$ (indicative of a constant density shell) and $s = 2$ (indicative of an $r^{-2}$ steady state wind).  Both fits yielded unphysically massive CSM compared to the ejecta mass, with $M_{\rm ej} = 9.17^{+2.56}_{-2.54}$ M$_{\odot}$ and M$_{\rm CSM} = 32.89^{+3.73}_{-3.20}$ M$_{\odot}$ for the $s = 0$ model and M$_{\rm ej} = 0.23^{+0.16}_{-0.07}$ M$_{\odot}$ and M$_{\rm CSM} = 39.85^{+6.50}_{-6.75}$ M$_{\odot}$ for the $s = 2$ model.  However, parameters from semi-analytic models, such as the one employed by MOSFiT, are known to be inconsistent with those found by numerical approaches \citep{2013MNRAS.428.1020M, SorokinaEtal2016, MoriyaEtal2018}, and can only be properly determined from non-LTE radiation hydrodynamical modeling \citep{ChatzopoulosEtal2013}.  Thus, even though our fit parameters were unphysical, we do not outright reject the CSM interaction scenario because of this.

\citet{Janka2012} finds that SN explosion models driven by neutrinos are unlikely to explain SN energies above $\sim2\times10^{51}$ erg. However, we estimate that the total kinetic energy of SN\,2020qlb is $\sim2\times10^{52}$ erg (see Sect.~\ref{Magnetar_MCMC}).  Since CSM interaction can only draw its energy from the kinetic energy, and given that the explosion mechanism in an ejecta-CSM interaction powered-SN is neutrino-driven, it is unlikely to be the main power source of SN\,2020qlb's light curve unless a magnetar is also present to supply the necessary additional energy.

One could expect X-rays from the shock created by the interaction, similar to what has been seen for Type IIn supernovae \citep[e.g.,][]{ChandraIIxray, KatsudaIIxray}. In soft X-rays, this emission should be dominated by line emission.  The strength of the emission also depends heavily on whether the shock is radiative or adiabatic, the CSM profile, the shock velocity, and other things for which we have no constraints.  Also, unlike Type IIn SNe, the shock should be surrounded by metal-rich CSM, which may absorb X-rays up to two orders of magnitude more efficiently.  From this, we find that an X-ray nondetection also seems consistent with the CSM interaction model. However, getting any meaningful constraints on physical parameters from our observed X-ray upper limit is unlikely.

Given the success of the magnetar model, the lack of spectral evidence for CSM interaction, unphysical fit parameters, energy considerations, as well as the large number of required CSM parameters we tentatively reject the CSM model as the primary light curve power source.

\subsection{Undulation source} \label{UndulationSource}

The magnetar model fits the general shape of SN\,2020qlb's light curve wherein undulating residuals remain. In this section we discuss possible mechanisms behind the observed modulation.

In Sect.~\ref{Undulations} we determine that SN\,2020qlb has more than two full periods of $32\pm6$ day undulations in the magnetar model residual near the peak of the bolometric light curve. The amplitude of the undulations was approximately $1.7\times10^{43}$ erg/s which is roughly 6\% of the peak bolometric luminosity. Another highly sampled SLSN-I, SN\,2015bn \cite[Fig. 24]{NichollEtal2016}, also had more than two full periods of magnetar residual undulations. SN\,2015bn, with a 30-50 day period oscillation amplitude of about $2.5\times10^{43}$ erg/s which is roughly 11\% of its peak bolometric luminosity, is similar to SN\,2020qlb with regard to its magnetar residual undulations. Intriguingly, both have two to three oscillations near peak brightness with timescales in the order of 30 days, and with roughly similar amplitudes.

In the following subsections we consider the source of the observed undulating magnetar residuals grouped into four possibilities: (1) variations in the centrally located power source, (2) variations in the SN ejecta properties, (3) interactions with varying CSM densities, or (4) the eventual breakdown of model assumptions.

\subsubsection{Central source fluctuations} \label{CentralSource}
Eventual central engine luminosity fluctuations will be stretched and delayed as they move through the homologous SN ejecta. The diffusion of photons through the ejecta thereby acts as a low-pass filter on any variable source. Central variations on short timescales compared to the timescale of the ejecta are therefore not expected to be observed at the SN photosphere.

One proposed central source for the undulations is suggested by \citet{Metzger2018} wherein fallback accretion onto the SN's central compact object could provide additional luminosity. The accretion rate is predicted to have a time dependence of $1/(1+t/t_{fb})^{5/3}$, where $t_{fb}$ is the fall-back timescale, which is different than the magnetar's luminosity time dependence (see Equation~\ref{eq:MagnetarPower}). 

A second possibility involves the eventual variability of the magnetar. \citet{DenissenyaEtal2021} discovered a local Milky Way magnetar (SGR1935+2154), the source of two fast radio bursts (FRBs), that showed a 231 day periodic windowed behavior (PWB) between epochs of activity and inactivity. Younger magnetars with significantly shorter spin periods could conceivably contain shorter periodic behavior as well, for example a 32 day pulsation.

\citet{ChugaiUtrobin2021} and \citet{MoriyaUnd} suggest a third possibility wherein a post-maximum enhancement of the central magnetar’s dipole field or the thermalization parameter (how much magnetar energy is converted into SN thermal energy) could cause a light curve bump. We note that the physical mechanism behind the enhancement is not yet known, that \citet{MoriyaUnd} predict an increase in photospheric temperature that is not detected in SN\,2020qlb, and that \citet{ChugaiUtrobin2021} only claim to explain a single bump. 

However it is difficult, if not impossible, for a pulsating central source on a scale of 32 days to diffuse through an ejecta with a diffusion timescale of 86 days (see Sec.~\ref{Magnetar_LeastSquares}). Following \citet[Eq. 8]{HosseinzadehEtal2021} (shown here in Equation~\ref{eq:RuleOfThumb}) we can constrain the depth from where a bump is produced. $\delta$-parameter values,
\begin{equation}\label{eq:RuleOfThumb}
\delta \equiv \dfrac{t_{bump}\times\Delta t_{bump}}{t_{rise}^{2}} < 1 ,
\end{equation}
rule out a central source of a light curve bump, where $t_{bump}$ is the time of the bump, $\Delta t$ is the bump duration (one period), and $t_{rise}$ is the rise time from explosion to peak.

When applying this rule of thumb to SN\,2020qlb's three residual maxima (at $t_{bump}$ $=$ 44, 76 \& 108 days post explosion) we calculate the $\delta$-parameter to be 0.24, 0.41 and 0.59. The fluctuation source(s) must therefore be well away from the center. We thereby, given the model assumptions, rule out any centrally located undulation source. See Section~\ref{Breakdown} for a discussion about possible assumption breakdowns.

\subsubsection{Ejecta property variations}
Fluctuations are seen in all filter bands as well as the bolometric light curve (see Figs. \ref{BoloLCinterpolated} \& \ref{UndulationsAllFilters}). No 30 day fluctuations are seen in the $g-r$ color evolution plot (Fig.~\ref{fig:grEvol}). These observations therefore suggest that there is no apparent wavelength dependence of the magnetar residual undulations.

\begin{figure}
        \centering
        \includegraphics[width=\columnwidth]{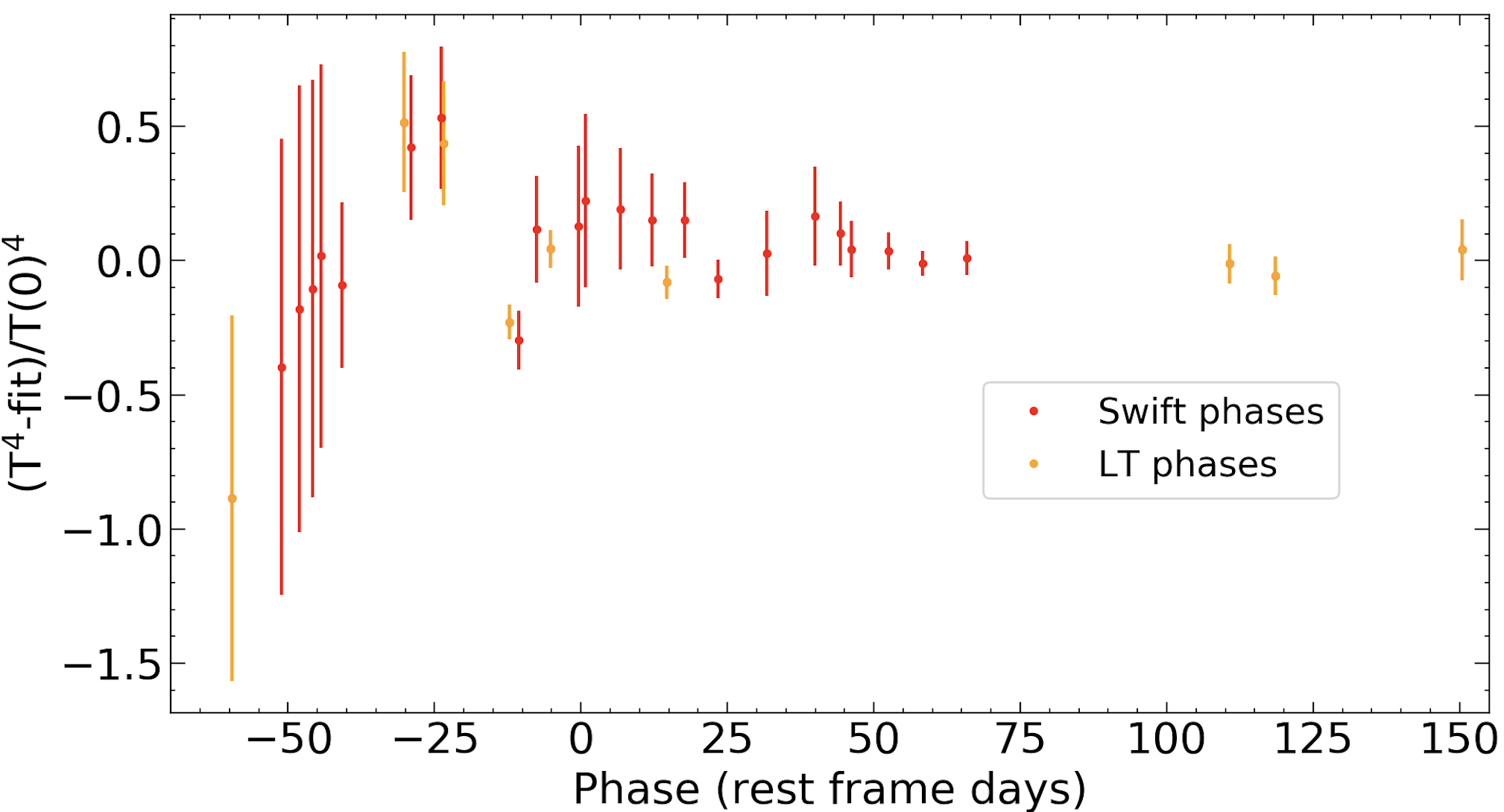}
        \caption{\label{TempVsMagnetar} The T$^{4}$ third degree polynomial residual relative to the zero phase T$^{4}$ in red at \textit{Swift} phases and in orange at LT phases.}
\end{figure}

In Fig.~\ref{TempVsMagnetar} we plot the residuals from the fourth power of the temperature (T$^{4}$) and a fitted third degree polynomial, relative to the fourth power of the zero phase temperature (11366 Kelvin) estimate. There are hints of small temperature bumps at -30, zero and +30 to +40 post peak days, including a three sigma drop between -23 and -12 days which matches a bolometric residual trough. However, within the uncertainties and in general, it is not clear that temperature changes match the bolometric light curve fluctuations.

\citet{KasenBildsten2010} hypothesize that magnetar winds could sweep up most of the ejecta into a dense shell with uniform velocity and a sharp temperature jump at the edge. The post peak receding photosphere would get hotter as it crosses this temperature jump adding luminosity to the light curve. This scenario could give credence to a single light curve bump (or a plateau), but not to the cyclic undulations as observed. We therefore rule out this temperature jump hypothesis for SN\,2020qlb.

\citet{MetzgerEtal2014} suggest that the magnetar wind nebula could inject electron/positron pairs into the base of the ejecta which would cool via Compton scattering and synchrotron emission. The resulting X-rays could ionize the inner portion of the SN ejecta forming ionization fronts which could propagate outward. Given the right conditions a front could break through the SN photosphere releasing unattenuated luminosity in both the optical/UV and soft X-ray bands. For instance, if an O\,II layer breaks through, the opacity to UV photons would be reduced. The additional leakage of UV photons through the photosphere could then disproportionately affect shorter wavelength UV observations. Given the lack of evidence for wavelength dependence of the undulations, and the nondetection of X-rays from 0.3-10 keV, we tentatively reject this hypothesis.

\citet{NichollEtal2016} note the possibility that central overpressure from a magnetar could drive a second shock wave through the expanding ejecta which could break through the SN photosphere at large radii. Estimates suggest that the effects of this mechanism should be comparably weak and occur typically within 20 days after explosion. Since this secondary shock wave hypothesis would also result in a single perturbation we rule it out for SN\,2020qlb.

Each of the above hypotheses essentially involves changes in the SN ejecta to create a light curve perturbation. In order to become more credible it will be necessary for such hypotheses to produce the general form of the observed undulations while also reproducing the observed spectral evolution. In the absence of further relevant evidence we disfavor this set of ejecta property undulation sources for SN\,2020qlb.

\subsubsection{External source fluctuations}\label{ExternalSource}

Ejecta interactions with density fluctuations in the CSM is the primary external source hypothesis to create undulations in the SN light curve. The open question in this subsection is therefore, what is(are) the mechanism(s) behind these eventual density fluctuations.

A collisional interaction between the SN ejecta and, for example, concentric spheres of CSM created from pre-explosion pulsational nuclear flashes from within a massive progenitor star could conceivably cause significant SN light curve undulations. \citet{Woosley2017} used hydrodynamic models of stars with M$_{ZAMS}$=70-140M$_{\odot}$ which typically end their lives as pulsational pair instability SNe (PPISNe). Magnetar power sources were included in the analysis. A broad range of possible outcomes was discovered wherein shells of CSM created by pulsational pair-instability (PPI) were found to have velocities in the range of $2000-4000$ km~s$^{-1}$, although with highly different kinetic energies and ejected masses. Fast moving SN ejecta could possibly catch up and interact with slower moving shells, depending on when they were ejected. In the right conditions, these precursors could even have luminosities similar to the peak of the supernovae \citep{Yoshidaetal16,Woosley2017}. At least one SLSN-I has been observed to have a circumstellar shell with a velocity of $\sim 3000$ km
~s$^{-1}$, consistent with a PPI origin. However, due to its large distance, the shell was seen through light echo scattering rather than direct interaction \citep{LunnanEtal2018b}.

To search for evidence of SN\,2020qlb precursor PPI events that were capable of ejecting a CSM shell we obtained a forced-photometry light curve \citep{YaoEtal2019} using all ZTF data since the beginning of the survey in March 2018. We apply quality cuts similar to \citet{StrotjohannEtal2021} and reject difference images that are flagged, have a seeing $>4\,\text{arcsec}$, have large residuals in the background region around the SN, or bad pixels at the SN position. In addition, we exclude any observations that are potentially affected by intermittent clouds\footnote{using the criteria described in Sect.~4.2 of \url{https://web.ipac.caltech.edu/staff/fmasci/ztf/extended_cautionary_notes.pdf}}. After quality cuts we are left with in total 1711 pre-explosion observations in the $g$-, $r$- and $i$-bands in 442 different nights.

\begin{figure}
        \centering
        \includegraphics[width=\columnwidth]{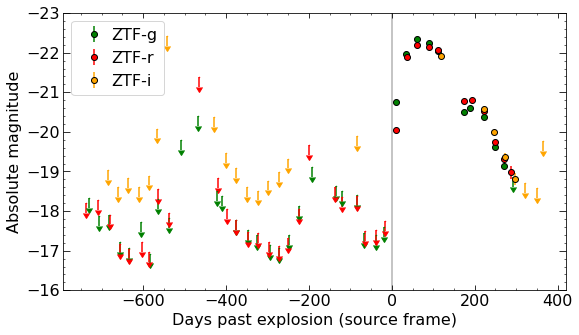}
        \caption{\label{fig:precusor_lc} ZTF forced-photometry light curve for SN\,2020qlb binned in 30-day-long bins. We did not detect any precursor outbursts and here show the 5\,$\sigma$ upper limits on pre-explosion outbursts. In the $g$- and $r$-bands, we typically obtain limiting magnitudes of $-18$ assuming that the outburst lasts for at least one month.}
\end{figure}

We do not detect any precursor events at the position of SN\,2020qlb when searching unbinned or binned (1 to 90-day-long bins) light curves following the methods described by \citet{StrotjohannEtal2021}. Absolute magnitude upper limits for 30-day-long bins are shown in Fig.~\ref{fig:precusor_lc}. In the $r$-band, the position was monitored in 27 out of 29 months within the 2.3 years before the SN explosion and we can rule out precursors brighter than magnitude $-18$ in 15 months, that is 52\% of the time. Precursors as bright as magnitude $-19$ would have been detected 83\% of the time, while precursors as faint as magnitude $-17$ would have remained undetected. \citet{Woosley2017} predict a wide range of possible PPISN precursor luminosities ($10^{41}$ to $10^{44}$ erg s$^{-1}$) over periods of weeks to millennia. The equivalent detection limit for magnitude $-18$ ($\approx5\times10^{42}$ erg s$^{-1}$) is therefore in the midst of expected PPI luminosities. Furthermore, \citet{SmithEtal2011} found that luminous blue variable (LBV) star eruptions had maximum brightnesses of $\gtrsim-15$ mag, all of which would have been undetected here. Moreover, low CSM shell velocities could place the time of the PPI event before the start of the ZTF survey. The PPI CSM mechanism is hence not definitely ruled out by the absence of bright precursor detections.

As discussed in SubSect.~\ref{CSMinteraction}, most CSM-interacting SNe have strong and narrow emission lines as encountered in the spectra of H-rich type IIn. No such spectral emission lines are seen herein for SN\,2020qlb or for the bump phases of many other SLSNe-I \citep{Nicholl2021}. A possible explanation might be the overwhelming luminosity of a SLSN near its peak compared to the weak emission luminosity of low masses of expanding PPI ejected shells of CSM. In addition, if the PPI ejecta have velocities of up to 4000 km~s$^{-1}$ the emission lines would not be narrow.

\citet{HosseinzadehEtal2021} estimate the typical mass of a CSM shell to be only $0.034^{+0.055}_{-0.027}M_{\odot}$ in order to power a bump. In contrast, a CSM powered SN would require a higher mass of CSM to interact with the SN ejecta which, in turn, could generate the characteristic strong and narrow emission lines. \citet{LiuEtal2018} (table 1) when modeling the complete light curves of SLSNe iPTF15esb and iPTF13dcc with only CSM interaction required a total of $4.09^{+0.42}_{-0.32}M_{\odot}$ and $25.34^{+4.67}_{-7.64}M_{\odot}$ to successfully model the two light curves. The lower CSM masses required to power a bump would therefore emit much weaker narrow emission lines, and presumably not be observed, than would the significantly higher CSM masses required to power an entire SLSN light curve.

External CSM density fluctuations could conceivably be caused by other pre-SN processes. One possible mechanism, known as wave-driven mass loss \citep{qs12, sq14}, involves super-Eddington fusion in the post-carbon burning phase, which could generate convection that could create acoustic waves capable of unbinding several solar masses of the stellar envelope in the last months or years prior to the SN explosion.

An additional possible source of CSM density fluctuations might be the interaction of a SN progenitor star's binary companion with an otherwise smooth CSM. \citet{SchwarzPringle1996} modeled the undulations in the radio light curve of SN\,1979C \citep{WeilerEtal1992} by a companion star's periastron passage using hydrodynamical simulations to determine that pronounced and asymmetrical spiral patterns in a massive (red supergiant) star's CSM can be formed. They also point out that a low viewing angle to the binary's orbital plane is important to observe the effects of the density variations. \citet{RyderEtal2004} discuss this mechanism as a possible explanation to modulations in the Type II SN\,2001ig radio band light curves. \citet{MorrisEtal2006} and \citet{MauronHuggins2006} discuss the Hubble Space Telescope (HST) image of LL Pegasi (AFGL 3068 or IRAS 23166+1655) as showing nested spiral shells of CSM predicted to occur when a mass losing star has a binary companion. \citet{MaerckerEtal2012} found a similar spiral pattern of CSM around R Sculptoris when using the Atacama Large Millimeter/submillimeter Array (ALMA). Others \citep[e.g.,][]{FraserEtal2013} mention this mechanism as a possible source of variable CSM interaction. This hypothesis is compelling as it could recreate the oscillatory form of the light curve residuals. However, the eventual robustness in the likelihood of its occurrence requires further statistical study.

The inspiral of a compact object (neutron star or a black hole) into the helium core of a massive binary companion star, which could expel stellar material in the form of a slowly expanding, dense and toroidally formed CSM prior to triggering conditions for a SN explosion, has been discussed by \citep[e.g.,][]{Chevalier2012} and modeled by \citet{SchroderEtal2020}. This type of merger-driven explosion provides a natural mechanism capable of creating aspherical CSM that could produce undulations in SLSNe-I light curves. Future efforts using multidimensional analysis of the explosion mechanisms and their resulting light curves are required to add credence to this compelling scenario.

The more general hypothesis of ejecta interaction with density fluctuations in the CSM as the undulation source must therefore be retained. As discussed, several differing processes could give rise to the required CSM density variations.

\subsubsection{Breakdown of assumptions}\label{Breakdown}
The idea that assumptions of simplified light curve models might break down already by the time of the SLSN peak luminosity or even earlier could conceivably explain unexpected phenomena. It is also possible to consider increased model intricacies to achieve the same end.

\citet{KaplanSoker2020} discuss how the sudden light curve drop observed in SN\,2018don could be modeled by jets driving the ejecta at the poles faster than at the equatorial regions. This scenario is expected to result in a strong initial light curve phase followed by an abrupt drop when the expansion eventually engulfs the asymmetry. The expectation here would be to find early time divergences from SN light curves otherwise unaffected by jets.

Spectropolarimetry has been used as a measure of the asymmetry of a SN which might be caused by, for example, failed jets or an asymmetric CSM \citep{Nicholl2021,WangWheeler2008}. Since magnetars have been suggested as gamma-ray burst (GRB) power sources, they could thereby launch jets, which if failed, could easily cause inner asphericities in the ejecta.  For instance, \citet{InserraEtal2016} found that SLSN-I SN\,2015bn showed significant polarization 24 days before and 27 days after maximum brightness, where the latter phase had a higher polarization. In addition, \citet{SaitoEtal2020} found that superluminous SN\,2017egm had higher polarization at late times. Both studies concluded that the inner ejecta were more aspherical than the outer ejecta. However, several other SLSNe have had detections consistent with zero polarization and thus have a spherical ejecta \citep[e.g.,][]{LeloudasEtal15, Cikotaetal18, Lee19, Lee20, Poidevin22}, implying that significant asymmetries may only appear in a minority of SLSNe.  

\citet{VurmMetzger2021} relaxed the assumption that a magnetar's power output is 100\% thermalized by the SLSN ejecta. Three dimensional simulations tracking the coupled evolutions of electron/positron pairs as well as photons in both the ejecta and the nebula were used to create a detailed model for the thermalization and escape of high energy radiation from the SN. The additional consideration of additional parameters and processes affecting gamma leakage and the effective opacity (normally assumed constant) could both put constraints on the magnetar parameters and add model flexibility.

Undulations could conceivably be caused by a geometric asymmetry of the ejecta. Spherical asymmetry could allow successive break-outs of hotter ejecta thereby brightening the light curve. For instance, the development of multi-dimensional hydrodynamical instabilities might create pockets wherein the effective diffusion time would be reduced, enabling energy from the central engine to flow relatively unimpeded through the ejecta brightening the light curve. Alternatively, successive blockages of brighter areas by regions of optically thick material could conceivably cause reductions in the SN's light curve. The continued use of 2D and 3D simulations which can reproduce anisotropies and hydrodynamical instabilities to more properly model the light curves of magnetar-powered SLSNe, for example as done by \citet{Chen16, SuzukiMaeda17, SuzukiMaeda21} (2D), \citet{SuzukiMaeda19, ChenKJetal2020} (3D), and \citet{BlondinChevalier2017} (2D and 3D), might bring clarity to this possibility.

\section{Conclusions} \label{Conclusions}
SN\,2020qlb is an extensively sampled SLSN-I that is among the most luminous (peak M$g$=-22.25 mags) and long-rising (72.4 days from 10$\%$ of maximum) SNe currently known. We estimate the total radiated energy of SN\,2002qlb to have been $\gtrsim2.8\pm{0.3}\times10^{51}$~erg. It exploded in a low-mass ($\log(M/M_{\odot}) = 7.5 _{-0.3}^{+0.6}$), low-metallicity ($12 + \log({\rm O/H}) = 8.0 \pm 0.2~{\rm dex}$, direct method) galaxy.

A large photometric data set is available and characterized by an excellent temporal coverage, even during the near peak solar conjunction by \textit{Swift}, broad wavelength coverage by the 14 measurement bands employed, and with a high measurement cadence by the ZTF survey telescope. We herein construct the SN\,2020qlb bolometric light curve, estimate the photospheric radius and temperature evolutions, and fit power source models to the resulting data.

We consider and rule out a $^{56}$Ni decay power source model due to unphysical parameter results. We disfavor the CSM power source model due to -- amongst other things -- the lack of the signature spectroscopic features of CSM interaction. We favor a model wherein the dipole spindown energy deposition of a rapidly rotating magnetar can power SN\,2020qlb's light curve. The magnetar model, using physically reasonable parameter values, results in a close fit to the majority of the bolometric and multi-band light curves.

During the first 150 days, the magnetar model residual has two to three oscillations with a 32$\pm$6 day timescale and an amplitude of about 6$\%$ of peak luminosity. Intriguingly, \citet{Nicholl2018} found a similar magnetar residual for the well-sampled SN\,2015bn where the near peak oscillatory form had a timescale of 30-50 days and an 11$\%$ amplitude.

We discuss three categories of hypotheses for the mechanism(s) behind the undulating magnetar residuals. A simple timescale argument rules out that a centrally sourced 32 day undulation could possibly survive the diffusion process as it passes through an ejecta with an 86 day diffusion timescale. In a second category, processes involving the ejecta are disfavored as they either predict a single bump or result in an unobserved wavelength dependence. We favor a third category, an external undulation source, which is the interaction of the SN ejecta with CSM density variations. Possible sources of the CSM density variations include pulsational pair instability, eruptive mass loss, the periastron passage of a companion star through an otherwise smooth CSM, and the common envelope evolution of a stellar merger-driven explosion. We also discuss whether possible breakdowns of model assumptions might result in light curve residuals.

To continue the research about the true source(s) of SLSNe undulations, we will need improved data breadth and resolution over time. An important upcoming project is the Vera Rubin Observatory's Legacy Survey of Space and Time (LSST) which is estimated to be able to discover $\sim10^{4}$ SLSNe per year with more than ten data points at redshifts up to $z\lesssim3$ \citep{VillarEtal2018}. Progress in modeling, data reduction and theory, large high cadence surveys such as ZTF, complemented by the depth of the LSST, should be capable of improving our understanding of the apparent SLSN undulations.

\begin{acknowledgements}
The authors would like to give special thanks to Matt Nicholl (Birmingham Univ.) for his helpful discussions regarding an early version of this work.

Some of the data presented herein were obtained at the W. M. Keck Observatory, which is operated as a scientific partnership among the California Institute of Technology, the University of California and the National Aeronautics and Space Administration. The Observatory was made possible by the generous financial support of the W. M. Keck Foundation.

Based on observations obtained with the Samuel Oschin Telescope 48-inch and the 60-inch Telescope at the Palomar Observatory as part of the Zwicky Transient Facility project. ZTF is supported by the National Science Foundation under Grants No. AST-1440341 and AST-2034437 and a collaboration including current partners Caltech, IPAC, the Weizmann Institute for Science, the Oskar Klein Center at Stockholm University, the University of Maryland, Deutsches Elektronen-Synchrotron and Humboldt University, the TANGO Consortium of Taiwan, the University of Wisconsin at Milwaukee, Trinity College Dublin, Lawrence Livermore National Laboratories, IN2P3, University of Warwick, Ruhr University Bochum, Northwestern University and former partners the University of Washington, Los Alamos National Laboratories, and Lawrence Berkeley National Laboratories. Operations are conducted by COO, IPAC, and UW.

The ZTF forced-photometry service was funded under the Heising-Simons Foundation grant nr. 12540303 (PI: Graham).

SED Machine is based upon work supported by the National Science Foundation under Grant No. 1106171.

The Liverpool Telescope is operated on the island of La Palma by Liverpool John Moores University in the Spanish Observatorio del Roque de los Muchachos of the Instituto de Astrofisica de Canarias with financial support from the UK Science and Technology Facilities Council.

This work has been supported by the research project grant “Understanding the Dynamic Universe” funded by the Knut and Alice Wallenberg Foundation under Dnr KAW 2018.0067, and the G.R.E.A.T research environment, funded by {\em Vetenskapsr\aa det}, the Swedish Research Council, project number 2016-06012.

Based on observations made with the Nordic Optical Telescope, owned in collaboration by the University of Turku and Aarhus University, and operated jointly by Aarhus University, the University of Turku and the University of Oslo, representing Denmark, Finland and Norway, the University of Iceland and Stockholm University at the Observatorio del Roque de los Muchachos, La Palma, Spain, of the Instituto de Astrofisica de Canarias.

The data presented here were obtained in part with ALFOSC, which is provided by the Instituto de Astrofisica de Andalucia (IAA) under a joint agreement with the University of Copenhagen and NOT.

This research is partially based on observations made with the NASA/ESA Hubble Space Telescope obtained from the Space Telescope Science Institute, which is operated by the Association of Universities for Research in Astronomy, Inc., under NASA contract NAS 5–26555. These observations are associated with program SNAP-16657 (PI Fremling).

This research has made use of the SVO Filter Profile Service (http://svo2.cab.inta-csic.es/theory/fps/) supported from the Spanish MINECO through grant AYA2017-84089.

T.-W. C. acknowledges the EU Funding under Marie Sk\l{}odowska-Curie grant H2020-MSCA-IF-2018-842471.

Steve Schulze acknowledges support from the G.R.E.A.T research environment, funded by {\em Vetenskapsr\aa det},  the Swedish Research Council, project number 2016-06012.

N. L. S. is funded by the Deutsche Forschungsgemeinschaft (DFG, German Research Foundation) via the Walter Benjamin program – 461903330.

\end{acknowledgements}

\bibliography{bibfile_BP} 
\bibliographystyle{aa} 

\end{document}